\documentclass[revtex4]{emulateapj}
\usepackage{graphicx}
\usepackage{lscape}
\usepackage{amssymb}
\usepackage{amsmath}
\usepackage{epstopdf}
\usepackage{footmisc}
\usepackage[flushleft]{threeparttable}
\usepackage{rotating}
\usepackage{longtable}
\begin{document}

\shorttitle{Distances and Ages of 22 Globular Clusters}
\shortauthors{O'Malley}

\title{Absolute Ages and Distances of 22 GCs using Monte Carlo Main-Sequence Fitting}

\author{Erin M. O'Malley}
\affil{Department of Physics and Astronomy\\
Dartmouth College, Hanover, NH 03784}
\email{Erin.M.O'Malley.GR@dartmouth.edu}

\author{Christina Gilligan}
\affil{Department of Physics and Astronomy\\
Dartmouth College, Hanover, NH 03784}

\author{Brian Chaboyer}
\affil{Department of Physics and Astronomy\\
Dartmouth College, Hanover, NH 03784}

\begin{abstract}
The recent \emph{Gaia} Data Release 1 of stellar parallaxes provides ample opportunity to find metal-poor main-sequence stars with precise parallaxes.  We select 21 such stars with parallax uncertainties better than $\sigma_\pi/\pi\leq0.10$ and accurate abundance determinations suitable for testing metal-poor stellar evolution models and determining the distance to Galactic globular clusters.  A Monte Carlo analysis was used, taking into account uncertainties in the model construction parameters, to generate stellar models and isochrones to fit to the calibration stars.  The isochrones which fit the calibration stars best were then used to determine the distances and ages of 22 globular clusters with metallicities ranging from -2.4 dex to -0.7 dex.  We find distances with an average uncertainty of 0.15 mag and absolute ages ranging from 10.8 -- 13.6 Gyr with an average uncertainty of 1.6 Gyr.  Using literature proper motion data we calculate orbits for the clusters finding six that reside within the Galactic disk/bulge while the rest are considered halo clusters.  We find no strong evidence for a relationship between age and Galactocentric distance, but we do find a decreasing age-[Fe/H] relation.

\end{abstract}


\section{Introduction}
Globular clusters (GCs) are among the oldest objects in our Galaxy that can be dated with a high level of precision; their ages provide a strict lower limit on the age of the Universe.  Although significant progress has been made in recent years, these estimates are still afflicted by uncertainty as high as $\sim1.5$ Gyr \citep{Grat1997,KC2003}.  The largest uncertainty stems from the dependence of GC distance determinations on the Population II distance scale with improvements possible only with more accurate parallax measurements.

The multitude of methods used to determine GC ages assume some comparison to stellar evolution models: horizontal branch morphology, white dwarf cooling sequence, or main-sequence turn-off (MSTO) location.  The technical details of each of these methods, along with their individual strengths and weaknesses, can be found in the review by \citet{SodRev2010}.  In general, the stellar evolution models, upon which these methods rely, are inherently uncertain due to uncertainties in the physical processes which occur in stars, such as incomplete equations of state and approximate treatment of convection by an arbitrary mixing length parameter, among others.

As the MS is the most well-understood phase of stellar evolution, MS-fitting can provide robust ages for GCs.  Calibration of this method at low metallicity requires accurate distances and abundances for very low metallicity stars.  The \emph{Hipparcos} catalog provides accurate parallaxes of over 115,000 stars that could be used to study any number of astrophysical problems.  However, only a few hundred of these stars are metal-poor dwarfs, of which only a handful are suitable for identifying the location of the MS in our stellar isochrones.  Nevertheless, when the Hipparcos catalog was first released to the community many groups attempted to use the parallax data available to calibrate stellar evolution models and study the distances and ages of GCs \citep[e.g.][]{Grat1997,Chab1998,Carr2000,Grun2002,Grat2003}.  The stringent constraints used in these studies (typically $\sigma_\pi/\pi<0.12$, M$_V>5.5$) limited the available sample to 10 -- 15 stars in most cases (with -1.6$<$[Fe/H]$<$-1.0), with the exception of a \citet{Grun2002} who were interested in a higher metallicity range ($-0.95<$[Fe/H]$<-0.65$) and therefore had a larger sample of about 20 stars and \citet{Grat1997}, \citet{Carr2000} and \citet{Grat2003} who extend their metallicity ranges to include higher metallicities and find samples of about 30 suitable stars for their calibrations.  

The studies with larger samples of calibration stars find GC distance uncertainties of $\sim0.06$ mag compared to those studies with fewer calibration stars which find GC distance uncertainties of $\sim0.1$ mag.  However, it should be noted that this improvement in GC distance estimates was not solely due to the increased number of calibration stars, but in large part to the use of improved and consistent metallicity scales and reddening estimates.  In general, these studies found that the \emph{Hipparcos} parallaxes were actually larger than the more uncertain ground-based measurements, leading to a determination of GC distance moduli that were on average 15\% larger than previously determined \citep{Grat2003} and smaller GC ages as a result.

Recently, \citet{Chab2016} present \emph{HST} parallaxes of eight metal-poor stars (with $-2.6<[Fe/H]<-1.6$) with uncertainties of $\sim1$\%.  The authors use updated stellar models and improved observations to provide robust distances and ages for nine very metal-poor GCs with the lowest overall uncertainty \citep{Chab2016}, ultimately demonstrating the improvements one can expect from the use of precise parallaxes of metal-poor stars.

In this paper, we use the methods of \citet{Chab2016} and present the results of MS-fitting distances and ages to 22 GC in the \emph{HST} ACS GC Treasury Project \citep{Sara2007} based on the recent Tycho-Gaia astrometric solution (TGAS) \citep{MLH2015,Gaia,Lindegren2016} parallaxes and the \citet{Chab2016} parallaxes for nearby subdwarfs with $-2.6\leq\mathrm{[Fe/H]}\leq-0.4$.  In $\S$\ref{Calibration} we present the sample of subdwarfs used in the calibration of our stellar evolution models and the results of this calibration.  We then present in $\S$\ref{Clusters} the GCs for which distances and ages will be determined using these calibrated stellar evolution models along with the weighted mean distance modulus and age for each cluster.  We discuss the significance of our findings in $\S$~\ref{Orbits} and $\S$~\ref{AgeRel} where we present updated orbital calculation for these clusters based on the newly derived cluster distances and literature proper motions along with age-galactocentric distance and age-[Fe/H] relations.  Lastly, we provide a summary of our work and future endeavors in $\S$\ref{Summary}.

\section{Calibration of Stellar Evolution Models with Local Subdwarfs}\label{Calibration}
In order to calibrate stellar evolution models at metallicities less than solar, low metallicity MS stars with precise parallax measurements and abundance determinations are needed, as the physical properties of MS stars are well-known and remain quite stable over time.  The recent release of the TGAS contains parallaxes for millions of stars with a sufficient number of metal-poor MS stars with precise parallaxes to allow a detailed calibration of metal-poor isochrones.

\subsection{Selection of Calibration Subdwarfs}
\citet{Chab2016} demonstrated the calibration of stellar evolution models is possible with the use of accurate parallaxes for metal-poor stars.  Here, we extend the range of metallicities over which a similar calibration can be performed by utilizing the TGAS parallaxes available as a result of the first \emph{Gaia} data release.  The first \emph{Gaia} data release includes stars from the Tycho-2 Catalog \citep{Tycho2} which provides positions and proper motions of the 2.5 million brightest stars. In order to select stars with reliable metallicities we cross-correlated the stars included in the Tycho-2 survey with the PASTEL catalog \citep{Pastel} of high resolution spectroscopy, finding 24,183 stars included in both catalogs.  Oftentimes, the PASTEL catalog includes multiple [Fe/H] measurements for a given star, including very old references; therefore, we take the star's [Fe/H] as the average of only those measurements more recent than the year 2000.  As the GCs we are examining are characteristically metal-poor, we are specifically looking to calibrate the location of the MS using metal-poor subdwarfs, [Fe/H]$\leq-0.5$ dex, a requirement that leaves us with 4,643 stars in the initial sample.

Among the criteria for selecting MS stars is the need for the star to be cool and have a high surface gravity.  In order to define the location of the MS we assume a MS magnitude of $M_V\geq5.5$ mag and use isochrones from the Dartmouth Stellar Evolution Database \citep[DSEP]{Dott2008} in a range of metallicities, $-3.0\leq\mathrm{[Fe/H]}\leq-0.5$ dex, [$\alpha$/Fe] = +0.4 dex and ages of both 10 and 13 Gyr to find the color of the MS as a function of metallicity.  We find the color of each isochrone at an absolute magnitude of $M_V = 5.5$ mag and find the best-fitting polynomial relation for V-I color as a function of metallicity given in Equation~\ref{eq:VIfit}.

\begin{multline}
\label{eq:VIfit}
V-I=0.895 + 0.258[\textrm{Fe/H}] + 0.104 [\textrm{Fe/H}]^{2} \\+ 0.015 [\textrm{Fe/H}]^{3}.
\end{multline}

As expected, we find this relation holds for both the 10 Gyr and 13 Gyr isochrones as the location of the MS is independent of age.

Initial UBVRI photometry were tabulated using SIMBAD and stars with observed V-I colors less than the theoretical color given by Equation \ref{eq:VIfit} were removed from our sample, leaving the remaining 415 stars as MS candidates.  Since our calibration stars need to have very accurate parameters in order to provide the best calibration of theoretical models, we remove from our sample the 29 stars that have photometry solely from photographic plates.

We estimated the reddening of our calibration stars by cross-correlating our sample with the Stromgren-Crawford uvby$\beta$ catalog \citep{SCcat}.  Stars that are within 80 pc of us do not exhibit much reddening \citep{Reis2011,Lallement2014} and are included in our final sample even if they do not have Stromgren photometry.  For stars with Stromgren photometry, we used the methods of both \citet{Schuster1989} and \citet{KS2010} to determine the reddening.  Both studies define the reddening, $E(b-y)=(b-y)-(b-y)_0$, as a function of three color indices: $m_1 = (v-b)-(b-y)$ used to measure line blanketing from metal lines, $c_1=(u-v)-(v-b)$ to measure the strength of the Balmer discontinuity, and $\beta = \beta_w-\beta_n$ which is sensitive to the Hydrogen $\beta$ line and stellar surface temperature.  For \citet{Schuster1989},

\begin{multline}
\label{eq:Schuster1989}
(b-y)_0 = 0.579 + 1.541m_0 - 1.066c_0 - 2.965(\Delta\beta)\\+9.64(\Delta\beta)^2(- 4.383m_0(\Delta\beta)- 3.821m_0c_0\\ + 6.695c_0(\Delta\beta) + 7.763m_0c_0^2.
\end{multline}

For \citet{KS2010},
\begin{multline}
\label{eq:Schuster1989}
(b-y)_0 = 0.492 - 0.976c_0 + 2.239(\Delta\beta) - 8.77(\Delta\beta)^2\\+ 6.26m_0c_0 - 16.15c_0(\Delta\beta)
-4.720m_0c_0^2 + 53.24c_0(\Delta\beta)^2\\+ 0.39(\Delta\beta)^2 + 27.526c_0^2(\Delta\beta) - 26.757m_0c_0(\Delta\beta).
\end{multline}

where, in both cases, $m_0 = m_1+0.3E(b-y)$, $c_0 = c_1-0.2E(b-y)$ and $\Delta\beta = 2.720 - \beta$.

We calculate $E(b-y)$ for our sample stars using both methods and find similar results for each; however, the \citet{KS2010} results gave lower residuals. Therefore, the reddening estimates were calculated using the newer method, except in cases where high resolution spectroscopy found no evidence for interstellar Na I lines \citep{OMprep}, implying that the reddening was negligible.  The reddening was converted from $E(b-y)$ to $E(V-I)$ using the relations of \citet{KS2010} and \citet{Wink1997}.

Finally, we test for the Lutz-Kelker bias \citep{LK1973} in our $\pi_{10}$ sample.  We provide a detailed description of our Lutz-Kelker analysis in the appendix for the interested reader, but ultimately find no appreciable correction in either absolute magnitude or metallicity.

Table~\ref{table:Stars} provides the observational stellar parameters for our final sample of 24 subdwarfs with accurate TGAS and/or \emph{HST} parallaxes ($\sigma_{\pi}/\pi\leq10\%$) along with references for the photometric and parallax measurements.  Three additional stars are included in this table that do not have TGAS or \emph{HST} parallaxes but which have accurate parallaxes determined by previous studies.  The first four columns give the star's ID, V magnitude, reddening listed as E(V-I) and the dereddened V-I color.  The last four columns give previously determined parallax measurements, the current TGAS parallaxes, the absolute magnitude and finally the metallicity of the star.  The absolute magnitudes are calculated using the TGAS parallaxes where available except for HIP\,87788, HIP\,54639, HIP\,46120 and HIP\,103269 which have \emph{HST} parallaxes that are both more accurate and more reliable than the TGAS parallaxes.

To directly show the impact the TGAS parallaxes have on our calibration of metal-poor stellar evolution models we construct two sub-sample groups: $\pi_{10}$ and $\pi_{12}$.  The $\pi_{10}$ group contains all of the stars in our sample that have either TGAS parallaxes, \emph{HST} parallaxes, or both with $\sigma_\pi/\pi\leq0.10$.  For the four stars that have both, we use the \emph{HST} parallaxes in the analysis as they have smaller uncertainties than the TGAS parallaxes.  The $\pi_{12}$ group contains all of the stars with previous \emph{Hipparcos} or \emph{HST} parallaxes with $\sigma_\pi/\pi<0.12$.

\begin{table*}[t]
\centering
\setlength{\tabcolsep}{10pt}
\begin{threeparttable}
\caption{Calibration Star Observational Properties \label{table:Stars}}
\begin{tabular}{l r c c r r c r}
\hline\hline
ID & V & E(V-I) & $(V-I)_{0}$ & $\pi_{prev}$(mas) & $\pi_{TGAS}$ (mas) & $M_{V}$ & [Fe/H]\\
\hline
HIP\,87788 & 11.30\tnote{a} & 0.00 & 0.85 & $10.83\pm0.13$\tnote{a} & $10.97\pm0.26$ & $6.47\pm0.04$ & -2.66\\
Wolf 1137 & 12.01\tnote{h} & 0.08 & 0.85 & $8.96\pm4.39$\tnote{i} & $7.56\pm0.28$ & $6.20\pm0.03$ & -2.53\\
HIP\,54639 & 11.38\tnote{a} & 0.00 & 0.91 & $11.12\pm0.11$\tnote{a} & $12.26\pm0.23$ & $6.61\pm0.04$ & -2.50\\
HIP\,106924 & 10.36\tnote{a} & 0.00 & 0.80 & $14.47\pm0.10$\tnote{a} && $6.16\pm0.04$ & -2.23\\
HIP\,46120 & 10.12\tnote{a} & 0.00 & 0.74 & $15.01\pm0.12$\tnote{a} & $14.94\pm0.21$ & $6.00\pm0.04$ & -2.22\\
HD\,321320 & 10.24\tnote{e} & 0.00 & 0.79 & $17.36\pm2.47$\tnote{i} & $16.65\pm0.25$ & $6.35\pm0.03$ & -1.98\\
HIP\,103269 & 10.27\tnote{a} & 0.00 & 0.77 & $14.12\pm0.13$\tnote{a} & $13.76\pm0.22$ & $6.02\pm0.05$ & -1.83\\
HIP\,108200 & 10.99\tnote{a} & 0.02 & 0.81 & $12.40\pm0.09$\tnote{a} && $6.41\pm0.04$ & -1.83\\
HD\,25329 & 8.50\tnote{d, f} & 0.05 & 1.22 & $54.12\pm1.08$\tnote{i} && $7.05\pm0.07$ & -1.80\\
HD\,188510 & 8.84\tnote{b, e, h} & 0.00 & 0.74 & $26.71\pm1.08\tnote{i}$ & $26.20\pm0.22$ & $5.93\pm0.03$ & -1.56\\
BD+511696 & 9.92\tnote{f, g} & 0.00 & 0.72 & $12.85\pm1.33$\tnote{i} & $13.93\pm0.25$ & $5.64\pm0.06$ & -1.51\\
HD\,134439 & 9.09\tnote{d, e} & 0.04 & 1.10 & $ 34.14\pm1.36$\tnote{i} && $6.66\pm0.07$ & -1.44\\
HD\,134440 & 9.46\tnote{d, e} & 0.06 & 1.20 & $33.68\pm1.67$\tnote{i} && $6.95\pm0.07$ & -1.42\\
HD\,97214 & 9.22\tnote{c} & 0.00 & 1.27 & $49.38\pm0.96$\tnote{i} & $50.46\pm0.40$ & $7.73\pm0.05$ & -1.38\\
HD\,145417 & 7.53\tnote{c, d, e} & 0.03 & 0.94 & $72.01\pm0.68$\tnote{i} & $73.65\pm0.30$ & $6.79\pm0.03$ & -1.27\\
Ross 484 & 10.84\tnote{d} & 0.00 & 1.17 & $28.8\pm2.9$\tnote{i} & $21.48\pm0.22$ & $7.42\pm0.07$ & -1.25\\
BD-033746 & 9.85\tnote{f} & 0.00 & 1.31 & $37.04\pm1.75$\tnote{i} & $38.87\pm0.23$ & $7.80\pm0.03$ & -1.22\\
CD-35360 & 10.24\tnote{d, e} & 0.00 & 0.85 & $15.49\pm1.54$\tnote{i} & $16.79\pm0.25$ & $6.37\pm0.03$ & -1.15\\
HD\,126681 & 9.29\tnote{d} & 0.00 & 0.72 & $21.04\pm1.12$\tnote{i} & $18.04\pm0.25$ & $5.57\pm0.03$ & -1.14\\
HD\,108564 & 9.46\tnote{d, e} & 0.00 & 1.14 & $36.78\pm1.01$\tnote{i} & $36.65\pm0.31$ & $7.32\pm0.07$ & -1.14\\
BD+080335 & 10.67\tnote{e} & 0.00 & 0.84 && $11.75\pm0.26$\tnote{i} & $6.01\pm0.07$ & -0.98\\
BD+022541 & 10.84\tnote{j} & 0.00 & 0.76 & $12.21\pm1.65$\tnote{i} & $12.39\pm0.23$ & $5.95\pm0.03$ &-0.88\\
HD\,230409 & 10.07\tnote{d, e} & 0.00 & 0.79 & $14.41\pm1.77$\tnote{i} & $14.51\pm0.24$ & $5.89\pm0.03$ & -0.86\\
HD\,092786 & 8.02\tnote{d, g} & 0.00 & 0.81 & $37.55\pm0.76$\tnote{i} & $36.97\pm0.30$ & $5.86\pm0.05$ & -0.81\\
HD\,144579 & 6.67\tnote{d, f} & 0.00 & 0.80 & $68.87\pm0.33$\tnote{i} & $69.56\pm0.22$ & $5.86\pm0.03$ & -0.68\\
HD\,073667 & 7.58\tnote{c} & 0.01 & 0.92 & $55.13\pm0.71$\tnote{i} & $54.00\pm0.33$ & $6.22\pm0.03$ & -0.59\\
HD\,216259 & 8.28\tnote{c, d} & 0.01 & 0.95 & $46.99\pm0.10$\tnote{i} & $44.81\pm0.30$ & $6.49\pm0.03$ & -0.59\\
\hline
\end{tabular}
\begin{tablenotes}
\footnotesize
\item [a] \citet{Chab2016}, [b] \citet{Marshall2007},[c] \citet{Koen2010}, [d] \citet{Cas10}, [e] \citet{Ryan1989}, [f] \citet{Ducati2002}, [g] \citet{Carney1987}, [h] \citet{Pancino2012}, [i] \citet{VL2007}, [j] \citet{ESA1997}
\end{tablenotes}
\end{threeparttable}
\vspace{0.25cm}
\end{table*}

\subsection{Comparison to Theoretical Isochrones}

We test the reliability of the DSEP isochrones \citep{Dott2008} using the same Monte Carlo analysis as in \citet{Chab2016}.  We will only briefly describe the process here.  The reliability of each model depends on the specific choices for the various physical parameters used to construct the stellar models; therefore, we vary the input parameters within probability distributions that are based on their intrinsic uncertainties.  Table~\ref{table:MCparams} provides the probability density distributions of the stellar evolution parameters that are varied in this study.  As an example, updated nuclear reaction rates of \citet{Adel2011} give \emph{pp-}chain reaction rates uncertain at the 1\% level.  In constructing the stellar evolution models for this study, the calculated nuclear reaction rate for the \emph{pp-}chain reaction is multiplied by a number drawn randomly from a Gaussian distribution with a mean and standard deviation of 1.00.  A total of 2000 independent isochrones are constructed in this manner to allow for the total theoretical uncertainty to be estimated via this MC analysis.

\begin{table*}[t]
\centering
\begin{threeparttable}
\caption{Monte Carlo Stellar Evolution Parameter Density Distributions \label{table:MCparams}}
\begin{tabular}{l c c c}
\hline\hline
Parameter & Distribution & Standard & Type\\
\hline
He mass fraction ($Y$)\dotfill & 0.24725 - 0.24757 & \citet{Planck2014} & Uniform\\
Mixing length\dotfill & 1.00 -- 1.70 ([Fe/H]$<-1.00$) & N/A & Uniform\\
& 1.20 -- 1.90 ([Fe/H]$\geq-1.00$) & N/A & Uniform \\
Convective overshoot\dotfill & $0.0H_p$ - $0.2H_p$ & N/A & Uniform\\
Atmospheric $T(\tau)$\dotfill & 33.3/33.3/33.3& \citet{Eddington} or & Trinary\\
&& \citet{Krishna} or &\\
&& \citet{Haus1999} &\\
Low-$T$ opacities\dotfill & 0.7 - 1.3 & \citet{Ferguson} & Uniform\\
High-$T$ opacities\dotfill & 1.00\% $\pm$ 3\% ($T \geq 10^7$ K) & \citet{IR1996} & Gaussian\\
Diffusion coefficients\dotfill & 0.5 - 1.3 & \citet{Thoul1994}& Uniform\\
$p + p\rightarrow \mathrm{H} + e^{+} + \nu_{e}^{2}$\dotfill& 1\% $\pm$ 1\% & \citet{Adel2011} & Gaussian\\
$^{3}\mathrm{He} + ^{3}\mathrm{He} \rightarrow ^{4} \mathrm{He} + 2p$\dotfill & 1\% $\pm$ 5\% & \citet{Adel2011} & Gaussian\\
$^{3}\mathrm{He} + ^{4} \mathrm{He} \rightarrow ^{7} \mathrm{Be} + \gamma$\dotfill &1\% $\pm$ 2\% & \citet{deBoer2014} & Gaussian\\
$^{12}\mathrm{C} + p \rightarrow ^{13}\mathrm{N} + \gamma$\dotfill & 1\% $\pm$ 36\% & \citet{Xu2013} & Gaussian\\
$^{13}\mathrm{C} + p \rightarrow ^{14}\mathrm{N} + \gamma$\dotfill & 1\% $\pm$ 15\% & \citet{Chak2015} & Gaussian\\
$^{14}\mathrm{N} + p \rightarrow ^{15}\mathrm{O} + \gamma$\dotfill & 1\% $\pm$ 7\% & \citet{Adel2011} & Gaussian\\
$^{16}\mathrm{O} + p \rightarrow ^{17}\mathrm{F} + \gamma$\dotfill & 1\% $\pm$ 16\% & \citet{Adel1998} & Gaussian\\
Triple-$\alpha$ reaction rate\dotfill & 1\% $\pm$ 15\% & \citet{Angulo1999} & Gaussian\\
Neutrino cooling rate\dotfill & 1\% $\pm$ 5\% & \citet{Haft1994} & Gaussian\\
Conductive opacities\dotfill & 1\% $\pm$ 20\% & \citet{HL1969} plus & Gaussian\\
&& \citet{Canuto1970}&\\
\hline
\end{tabular}
\begin{tablenotes}
\item {NOTE - As in \citet{BC2006}, parameters below atmospheric $T(\tau)$ are treated as multiplicative factors applied to standard tables and formulas.}
\end{tablenotes}
\end{threeparttable}
\end{table*}

Our choice of mixing length has been updated based on recent studies of metal-poor stars.  A uniform probability density distribution from 1.0 -- 1.7 was chosen for the mixing length in \citet{Chab2016} based on recent studies by \citet{Bonaca2012}, \citet{Tanner2014} and \citet{Creevey2015} which show lower metallicity stars requiring a mixing length parameter much lower than the solar mixing length $(\alpha_\odot=1.9)$.  In this paper the range of subdwarf and GC metallicities is extended towards higher metallicities; therefore, for models with [Fe/H]$>-1.00$ dex we shift the mixing length density distribution closer to solar, sampling a uniform density distribution from 1.2 -- 1.9.

For the majority of the subdwarfs in the sample we construct a suite of theoretical isochrones with a binary [$\alpha$/Fe] of $+0.2$ or $+0.4$ dex based on the findings of \citet[and references therein]{Sneden2004} that both metal-poor field stars and GC stars show Ca overabundance of +0.2 -- +0.4 dex for metallicities ranging from [Fe/H] = -2.4 dex to [Fe/H] = -0.8 dex.  For the most metal-poor star, HIP\,87788, \citet{OMprep} find [$\alpha$/Fe]$>+0.6$ and so for this star we construct a separate suite of models with a distribution of [$\alpha$/Fe] as follows: 0.40 dex (25\%), 0.60 dex (50\%), 0.80 dex (25\%).

The stellar evolution tracks, which cover a stellar mass range of 0.3 -- 1.0 M$_\odot$, are converted into isochrones using both the color-temperature relations of \citet{Haus1999} (ISO-P) and \citet{VC2003} (ISO-VC).  The goodness of fit of these isochrones is then determined using a reduced $\chi^2$ analysis with N degrees of freedom, where N is the number of stars in a given metallicity group.  It is well-known that the location of the zero-age main sequence is expected to be a function of metallicity based on theoretical stellar evolution models.  In this study we do not attempt to define this location as a function of metallicity, but instead perform our calibration on groups of subdwarfs spanning a narrow metallicity range of $\sim0.4$\,dex.

The median deviation of the isochrones from the $\pi_{10}$ calibration stars is $1.86\sigma$ for ISO-P and $1.92\sigma$ for ISO-VC, with the median of the individual metallicity groups ranging from 0.63 -- 3.65$\sigma$.  Table~\ref{MedSig} provides the average metallicity of each bin, the corresponding median standard deviations determined for both the ISO-P and ISO-VC models and the number of stars used in the determination.  The median deviation of the isochrones from the $\pi_{12}$ calibration stars is similar to or slightly smaller than that of the $\pi_{10}$ sample for four out of the six metallicity.  For the two metallicity bins in which the $\pi_{10}$ group uncertainty is smaller than that of the $\pi_{12}$ group, we attribute the larger uncertainty to the fact that although the parallax uncertainty in the $\pi_{12}$ group is allowed extend to $\sim12\%$, the average uncertainty is 3.6\% which is only slightly higher than that of the $\pi_{10}$ with an average uncertainty of 2.5\%.  Additionally, we find that the largest uncertainties stem from the three stars without TGAS or \emph{HST} parallaxes.  

\begin{figure*}[t]
\centering
\includegraphics[width=1.0\textwidth]{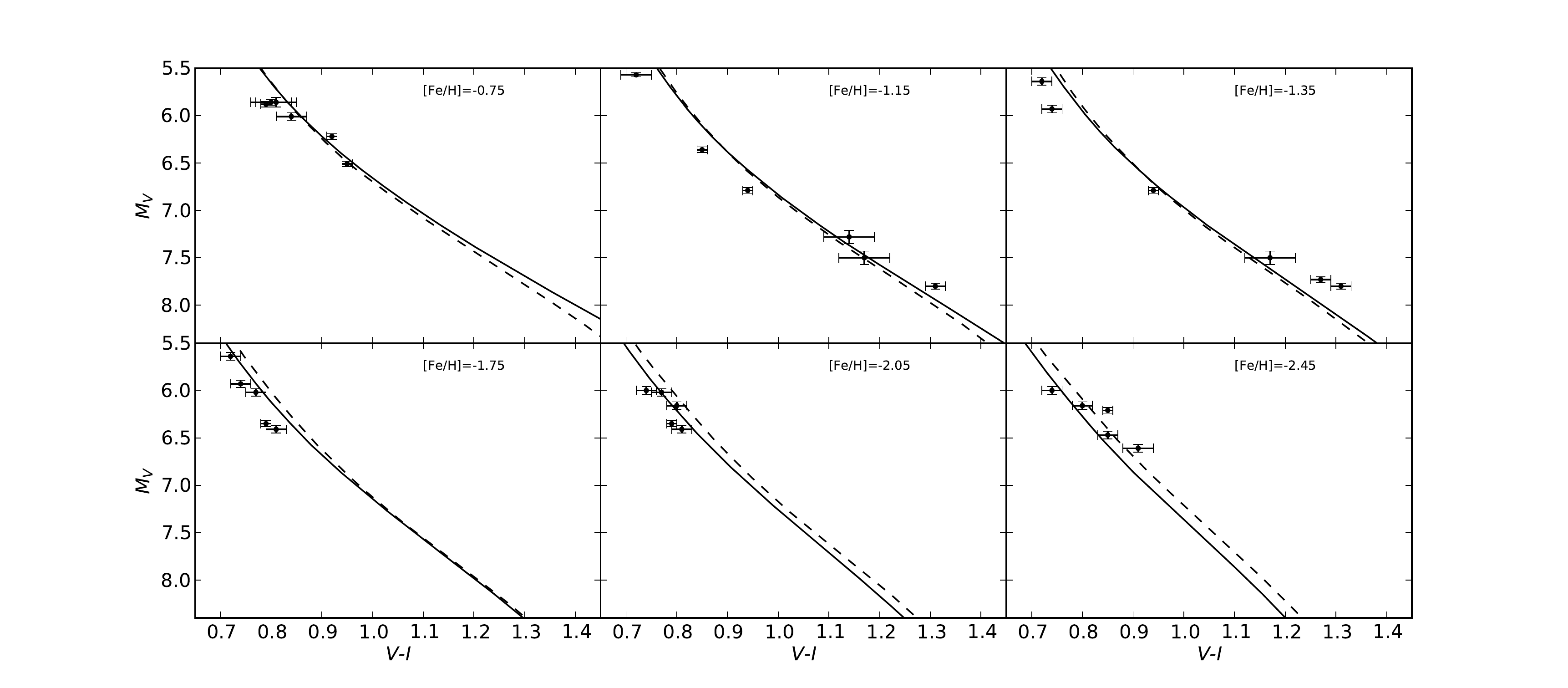}
\caption{\small{Comparison of $\pi_{10}$ calibration star locations in a CMD to Dartmouth stellar evolution isochrones of mean group metallicity.  Shown here are the ISO-P isochrones (dashed) and ISO-VC (solid) isochrones.  A distinct offset between the ISO-P and ISO-VC isochrones can be see in the most metal-poor groups.}}
\label{fig:isocomp}
\end{figure*}

We show in Figure~\ref{fig:isocomp} the location of the $\pi_{10}$ calibration stars in a color-magnitude diagram (CMD) compared to the average metallicity isochrone for their respective group, where the median ISO-P isochrone is shown as a dashed line and the median ISO-VC isochrone is solid.  From this figure one can see a noticeable offset between the ISO-P and ISO-VC isochrones in the most metal-poor regimes.  It is expected then that each set of isochrones in these regimes may give a slightly different distance modulus for a GC, but that weighting the GC distances based on the goodness-of-fit of the isochrones to the calibrations stars will allow us to find the true distance modulus of the GC.

\begin{table*}[t]
\centering
\caption{Median Standard Deviation of Isochrone Fits to Calibration Stars \label{MedSig}}
\begin{tabular}{l c c c | l c c c}
\hline\hline
\multicolumn{4}{c}{$\pi_{10}$}\vline & \multicolumn{4}{c}{$\pi_{12}$}\\
\cline{1-4}\cline{5-8}
[Fe/H] & N & $\sigma_P$ & $\sigma_{VC}$ & [Fe/H] & N & $\sigma_P$ & $\sigma_{VC}$ \\
\hline
-0.75 & 6 & 0.74 & 0.73 & -0.71 & 6 & 0.91 & 0.97\\
-1.20 & 6 & 3.10 & 3.14 & -1.20 & 6 & 2.21 & 2.40\\
-1.37 & 6 & 3.57 & 3.63 & -1.46 & 5 & 4.75 & 4.47\\
-1.74 & 5 & 0.95 & 0.75 & -1.68 & 5 & 1.46 & 1.14\\
-2.02 & 5 & 1.10 & 0.75 & -2.02 & 4 & 1.19 & 0.83\\
-2.43 & 5 & 1.80 & 2.63 & -2.40 & 4 & 1.23 & 2.25\\
\hline
\end{tabular}
\vspace{0.25cm}
\end{table*}

\begin{figure*}[t]
\centering
\includegraphics[width=1.0\textwidth]{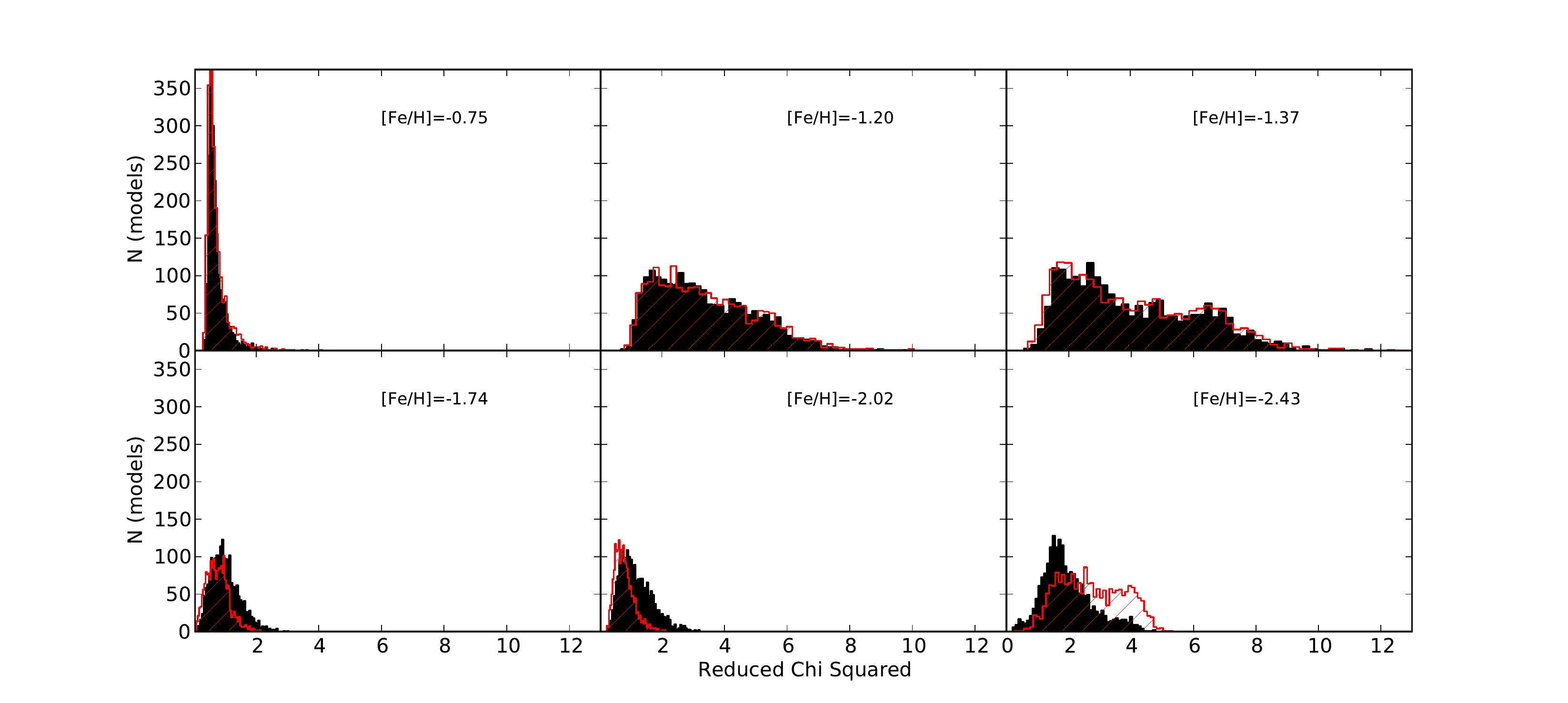}
\caption{The distribution of $\chi^2_\mathrm{red}$ is shown for both ISO-P (black) and ISO-VC (red dashed) in each metallicity bin.  The isochrones fit well the most metal-rich stars within a narrow range of $\chi^2_\mathrm{red}$ values.  The distribution of goodness of fit is more broad for the other metallicity bins.  In the most metal-poor bin, there is a noticeable difference in the goodness of fit produced by ISO-P versus ISO-VC.}
\label{fig:chihist}
\vspace{0.25cm}
\end{figure*}

Even within a given isochrone set, either ISO-P or ISO-VC, different stellar model construction parameters will produce isochrones with varying degrees of goodness of fit.  In Figure~\ref{fig:chihist}, we show the distribution of reduced $\chi^2$ values for the isochrones in each metallicity bin for both the ISO-P (black) and ISO-VC (red) sets of isochrones.  As expected, the isochrones in the most metal-rich bin are clustered toward small $\chi^2_{\mathrm{red}}$ as stellar evolution models already do a good job reproducing observations in this metallicity regime.  However, this is not the case for the other five metallicity bins in which a larger spread in the goodness of fit is readily noticeable.  Another obvious feature highlighted in the most metal-poor bin is difference in goodness of fits produced by the ISO-P versus the ISO-VC isochrones.

\section{Distance and Age Determination of Milky Way GCs}\label{Clusters}
\subsection{Selection of Sample Clusters}
The GCs used in \citet{Chab2016} were limited to those from the \emph{HST} ACS GC Treasury Project \citep{Sara2007} with [Fe/H] $<-1.4$ dex and E(B-V) $<0.10$ from the \citet{Harris1996} GC catalog.  We expand upon that study with the inclusion of additional clusters from the \emph{HST} ACS GC Treasury Project \citep{Sara2007} with [Fe/H]$<-0.6$ dex and E(B-V)$\leq0.15$ mag.

It is well known that the selection of color filters is important in GC studies as the effects of multiple stellar populations are prominent in both B-V and B-I where the blue filter is particularly sensitive to light element abundance variations and the long color baseline of B-I is capable of discerning He enhanced populations \citep{Piotto2007,Marino2008,Sbordone2011}.  Therefore, it was expected that our use of V and I filters would be advantageous in determining the distances and ages to GCs by minimizing the effects of multiple stellar populations.  Although this is still true in general, recent studies have shown that the He enhanced populations can be identified in V,V-I CMDs in some GCs \citep{Milone47Tuc,Milone6397,Milone6752,Milone7089,Milone2808} and that multiple sub-giant branches may be apparent \citep{Milone47Tuc,Piotto2012}.  

We cross-reference the ACS GCs that fit our criteria with the $m_{275}-m_{814},m_{275}$ CMDs of \citet{Piotto2015} to find a suitable sample of clusters for this study.  If multiple stellar populations are not present in the long color baseline of $m_{275}-m_{814}$, then it is unlikely for them to be visible in V, V-I allowing us to confidently choose the cluster for our analysis.  On our inspection of the \citet{Piotto2015} CMDs we remove NGC\,362, NGC\,1261, NGC\,1851, NGC\,6715, NGC\,6934 and NGC\,7089 due to noticeable multiple populations along the MS or SGB.  We are encouraged by the agreement with \citet{Piotto2012} which finds multiple distinct SGBs in V-I (F606W-F814W) in 47 Tuc and NGC\,1851 as well as broadened SGBs in NGC\,362, NGC\,6715 and NGC\, 7089.  We also remove from our ACS sample those clusters that do not have CMDs provided by \citet{Piotto2015}: NGC\,4147, NGC\,5139, NGC\,7006, and ARP\,2.  Our final sample contains 22 GCs.

Three clusters in our original ACS sample (47 Tuc, NGC 6752, and NGC 7089) have published population percentages, helium enhancements, and MS ridgelines in F606W-F814W.  Thus, it is possible to use the published data to determine the location of the primordial MS for comparison to our isochrones.
\begin{itemize}
\item \emph{47 Tuc} -- \citet{Milone47Tuc} use $m_{F275W}$ and $m_{F336W}$ data to identify two populations of stars on the MS.  They find the primordial population makes up only 18\% of the stars, the second population is enhanced in helium by $\Delta\mathrm{Y} = 0.04$ dex, and they provide MS ridgelines in $m_{F606W}-m_{F814W},m_{F814W}$ for both populations.  We find a zero-point offset between our CMD in $m_{F606W}-m_{F814W},m_{F606W}$ and that of \citet{Milone47Tuc}; however, the data in \citet{Anderson2009} agrees with that of \citet{Milone47Tuc}.  We find the difference between the primordial ridgeline in \citet{Milone47Tuc} and the median ridgeline in \citet{Anderson2009} to be -0.05 mag in F606W at MS colors of $0.60\leq m_{F606W}-m_{F814W}\leq0.67$.  We use the \citet{Sirianni2005} relations to transform to an offset of -0.06 mag in V which will be applied to our distance modulus calculation for 47 Tuc.  \citet{Piotto2012} find two SGBs in 47 Tuc offset by $\sim0.05$ mag in $m_{F814W}$ which we transform to 0.09 mag in V via the \citet{Sirianni2005} relations.  We will apply this to the SGB location of 47 Tuc in determining the age of the cluster.
\item \emph{NG 6752} -- \citet{Milone6752} find three stellar populations in NGC\,6752.  Pop\,A is primordial and contains 25\% of the stars, Pop\,B is enhanced in helium by $\Delta\mathrm{Y} = 0.01$ dex and contains 45\% of the stars, and Pop\,C is enhanced in helium by $\Delta\mathrm{Y}=0.03$ dex.  Like \citet{Milone47Tuc}, $m_{F275W}$ and $m_{F336W}$ data are used to identify the different populations and MS ridgelines in $m_{F606W}-m_{F814W},m_{F814W}$ are provided as well.  They find no distinction between the Pop\,A and Pop\,B ridgelines while there is an offset from Pop\,C.  However, because Pop\,A and Pop\,B make up 70\% of the stars, our median MS ridgeline should follow this combined population along the MS.
\item \emph{NGC 7089} -- \citet{Milone7089} find seven stellar populations in NGC\,7089 using $m_{F275W}$ and $m_{F814W}$ data with helium enhancements of up to $\Delta\mathrm{Y}=0.07$ dex.  The authors provide ridgelines of the primordial population along with combinations of [$\alpha$/Fe] enhancements and helium enhancements up to 0.4 dex and 0.084 dex, respectively.  Additionally, \citet{Piotto2012} show a spread of 0.15 mag in the SGB of NGC\,7089.  Although MS ridgelines are available for NGC\,7089 like 47 Tuc and NGC\,6752, we conservatively remove NGC\,7089 from our sample due to the highly complex nature of the multiple stellar populations.
\end{itemize}

We separate the full set of 22 clusters into 6 metallicity groups (-2.40$\leq$[Fe/H]$\leq$-2.30; -2.30$<$[Fe/H]$\leq$-1.80; -1.80$<$[Fe/H]$\leq$-1.50; -1.50$<$[Fe/H]$\leq$-1.20; -1.20$<$[Fe/H]$\leq$-0.90; -0.90$<$[Fe/H]$\leq$-0.60)  so we can perform MS-fitting using calibration stars covering narrow metallicity ranges.  We construct models using the same input parameters as those used in the comparison to the calibration subdwarfs but with [Fe/H], [$\alpha$/Fe], and helium abundance sampled in the appropriate distribution for GCs.  The galactic coordinates and Galactocentric distance from the \citet{Harris1996} GC catalog for each cluster along with \citet{Carr2009} metallicities and \citet{DB2000} far-infrared reddening values are provided in Table~\ref{table:Clusters}.

\begin{table*}[t]
\centering
\caption{MS-fitting High Metallicity Cluster Data \label{table:Clusters}}
\begin{tabular}{l l r r r c c}
\hline\hline
Cluster ID & Name & l($^\circ$) & b($^\circ$) & R$_\mathrm{GC}$ & [Fe/H] & E(B-V)$_{FIR}$\\
\hline
NGC 104 & 47 Tuc & 305.89 & -44.89 & 7.4 & -0.76 & 0.03\\ 
NGC 288 & & 152.30 & -89.38 & 12.0 &-1.32 & 0.01\\
NGC 2298 & & 245.63 & -16.00 & 15.8 & -1.96 & 0.22\\
NGC 4590 & M 68 & 299.63 & 36.05 & 10.2 & -2.27 & 0.06\\
NGC 5024 & M 53 & 332.96 & 79.76 & 18.4 & -2.06 & 0.03\\
NGC 5053 && 335.70 & 78.95 & 17.8 & -2.30 & 0.02\\
NGC 5272 & M 3 & 42.22 & 78.71 & 12.0 & -1.50 & 0.01\\
NGC 5466 && 42.15 & 73.59 & 16.3 & -2.31 & 0.02\\
NGC 5904 & M 5 & 3.86 & 46.80 & 6.2 & -1.33 & 0.04\\
NGC 6101 && 317.74 & -15.82 & 11.2 & -1.98 & 0.10\\
NGC 6205 & M 13 & 59.01 & 40.91 & 8.4 & -1.58 & 0.02\\
NGC 6341 & M 92 & 68.34 & 34.86 & 9.6 & -2.35 & 0.02\\
NGC 6362 & & 325.55 & -17.57 & 5.1 & -1.07 & 0.07\\
NGC 6541 & & 349.29 & -11.19 & 2.1 & -1.82 & 0.16\\
NGC 6584 & & 342.14 & -16.41 & 7.0 & -1.50 & 0.11\\
NGC 6652 & & 1.53 & -11.38 & 2.7 & - 0.76 & 0.11\\
NGC 6681 & M 70 & 2.85 & -12.51 & 2.2 & -1.62 & 0.11\\
NGC 6723 & & 0.07 & -17.30 & 2.6 & -1.10 & 0.16\\
NGC 6752 & & 336.49 & -25.63 & 5.2 & -1.55 & 0.06\\
NGC 6809 & M 55 & 8.79 & -23.27 & 3.9 & -1.93 & 0.14\\
NGC 7078 & M 15 & 65.01 & -27.31 & 10.4 & -2.33 & 0.11\\
NGC 7099 & M 30 & 27.18 & -46.84 & 7.1 & -2.33 & 0.05\\
\hline
\end{tabular}
\end{table*}

It was found in \citet{Chab2016} that \citet{DB2000} provide more accurate estimates of cluster reddening than \citet{Harris1996} based on the red giant branch color.  The GCs in this study are sufficiently far away that the use of the \citet{DB2000} far-infrared reddening values can be used with the reddening typically being enhanced over \citet{Harris1996} by 0.01 to 0.02 mag, but in some cases by as much as 0.08 mag.  These reddening values are converted to E(V-I) in this analysis with the \citet{Wink1997} relations.

\subsection{MS-Fitting Distances and Ages}
We perform MS-fitting on \emph{HST} photometry of each cluster, converted into V and I magnitudes via the \citet{Sirianni2005} transformations, and find distances to the clusters given by a fit of the each theoretical isochrone to the median MS ridgeline.  We improve upon the method used in \citet{Chab2016} by defining the median ridgeline using the more robust method of rotated histograms described in \citet{MF2009}.  

It was established in \citet{Chab2016} that equally well-fitting distance moduli were obtained by shifting the median ridgeline in both color and magnitude.  Therefore, the cluster distance modulus may be calculated using the average reddening and the uncertainty may be propagated using standard techniques. Given that $A_V = 3.1\times E(B-V)$, the reddening uncertainty used in this study, $\sigma_\mathrm{E(B-V)}\pm0.01$ mag corresponds to $\pm0.03$ mag uncertainty in the distance modulus. 

For a given set of isochrones (ISO-P or ISO-VC) the distance modulus typically spans a range of 0.6 mag.  However, not all isochrones fit the calibrating parallax stars with the same level of accuracy; therefore, the distance modulus for each cluster is weighted based on the goodness of fit of the isochrone to the calibration stars.

We use the distance modulus for each cluster and isochrone combination to find the absolute magnitude of the cluster SGB which we compare to the SGB magnitude of isochrones ranging in age from 8 -- 15 Gyr.  \citet{Chab1996} demonstrate that the SGB magnitude at the location 0.05 mag redder and more luminous than the MS turn-off is an excellent age indicator which minimizes the uncertainty in the derived ages.  We apply the same weighting scheme to the ages as we did to the distances to derive ages that incorporate the goodness of fit measure to the calibration stars.  The age for each GC is provided along with the distance modulus in Tables~\ref{table:Results} for each isochrone set, ISO-P and ISO-VC, along with final, combined results.

\begin{table*}[t]
\centering
\setlength{\tabcolsep}{4pt}
\caption{$\pi_{10}$ Weighted GC Distance Modulus and Age \label{table:Results}}
\begin{tabular}{rrrrrrrrr}
\hline\hline
& \multicolumn{2}{c}{ISO-P} & & \multicolumn{2}{c}{ISO-VC} & & \multicolumn{2}{c}{Combined}\\
\cline{2-3}\cline{5-6}\cline{8-9}
Cluster & $(m-M)_V$ & Age (Gyr) && $(m-M)_V$ & Age (Gyr) && $(m-M)_V$ & Age (Gyr)\\
\hline
NGC 104 & $13.57\pm0.20$ & $11.5\pm2.0$ && $13.56\pm0.20$ & $11.6\pm2.1$ && $13.56\pm0.20$ & $11.6\pm2.0$\\
NGC 288 & $14.91\pm0.15$ & $11.4\pm2.0$ && $14.90\pm0.15$ & $11.6\pm2.0$ && $14.91\pm0.15$ & $11.5\pm2.0$\\
NGC 2298 & $15.65\pm0.12$ & $12.1\pm1.4$ && $15.55\pm0.11$ & $13.6\pm1.5$ && $15.61\pm0.11$ & $12.9\pm1.5$\\
NGC 4590 & $15.36\pm0.10$ & $12.7\pm1.2$ && $15.40\pm0.10$ & $12.4\pm1.3$ && $15.38\pm0.10$ & $12.5\pm1.3$\\
NGC 5024 & $16.56\pm0.11$ & $13.3\pm1.3$ && $16.59\pm0.11$ & $13.0\pm1.3$ && $16.58\pm0.11$ & $13.1\pm1.3$\\
NGC 5053 & $16.29\pm0.11$ & $13.3\pm1.2$ && $16.32\pm0.10$ & $13.1\pm1.3$ && $16.31\pm0.10$ & $13.2\pm1.3$\\
NGC 5272 & $15.19\pm0.13$ & $11.2\pm1.7$ && $15.17\pm0.14$ & $11.4\pm1.8$ && $15.18\pm0.14$ & $11.3\pm1.8$\\
NGC 5466 & $16.15\pm0.11$ & $13.5\pm1.2$ && $16.19\pm0.10$ & $13.2\pm1.3$ && $16.16\pm0.10$ & $13.4\pm1.3$\\
NGC 5904 & $14.58\pm0.15$ & $10.7\pm2.0$ && $14.56\pm0.15$ & $10.9\pm2.0$ && $14.57\pm0.15$ & $10.8\pm2.0$\\
NGC 6101 & $16.03\pm0.11$ & $13.7\pm1.3$ && $16.07\pm0.11$ & $13.5\pm1.4$ && $16.06\pm0.11$ & $13.6\pm1.5$\\
NGC 6205 & $14.54\pm0.12$ & $12.3\pm1.5$ && $14.55\pm0.13$ & $12.1\pm1.7$ && $14.55\pm0.13$ & $12.2\pm1.7$\\
NGC 6341 & $14.85\pm0.11$ & $13.0\pm1.2$ && $14.83\pm0.12$ & $13.3\pm1.3$ && $14.84\pm0.12$ & $13.1\pm1.5$\\
NGC 6362 & $14.76\pm0.19$ & $11.3\pm2.0$ && $14.76\pm0.20$ & $11.4\pm2.1$ && $14.76\pm0.19$ & $11.4\pm2.0$\\
NGC 6541 & $14.98\pm0.12$ & $12.6\pm1.4$ && $15.02\pm0.12$ & $12.3\pm1.5$ && $15.00\pm0.12$ & $12.4\pm1.5$\\
NGC 6584 & $16.16\pm0.13$ & $11.6\pm1.5$ && $16.16\pm0.13$ & $11.7\pm1.7$ && $16.16\pm0.13$ & $11.6\pm1.7$\\
NGC 6652 & $15.28\pm0.19$ & $11.4\pm2.0$ && $15.27\pm0.19$ & $11.5\pm2.1$ && $15.27\pm0.19$ & $11.4\pm2.0$\\
NGC 6681 & $15.26\pm0.13$ & $12.7\pm1.5$ && $15.28\pm0.13$ & $12.6\pm1.7$ && $15.27\pm0.13$ & $12.7\pm1.7$\\
NGC 6723 & $14.87\pm0.20$ & $11.9\pm2.0$ && $14.86\pm0.20$ & $12.0\pm2.1$ && $14.86\pm0.20$ & $11.9\pm2.0$\\
NGC 6752 & $13.36\pm0.13$ & $12.7\pm1.6$ && $13.37\pm0.13$ & $12.6\pm1.7$ && $13.37\pm0.13$ & $12.6\pm1.7$\\
NGC 6809 & $14.17\pm0.12$ & $11.6\pm1.3$ && $14.20\pm0.11$ & $11.3\pm1.4$ && $14.19\pm0.11$ & $11.4\pm1.4$\\
NGC 7078 & $15.68\pm0.11$ & $12.7\pm1.2$ && $15.71\pm0.10$ & $12.4\pm1.3$ && $15.69\pm0.10$ & $12.6\pm1.3$\\
NGC 7099 & $14.87\pm0.11$ & $12.8\pm1.2$ && $14.89\pm0.10$ & $12.5\pm1.3$ && $14.88\pm0.11$ & $12.7\pm1.3$\\
\hline
\end{tabular}
\vspace{0.25cm}
\end{table*}

One might be interested in the improvements gained using the $\pi_{10}$ sample of calibration stars versus the $\pi_{12}$ sample.  We perform the same MS-fitting and age dating analysis weighting the resulting distances and ages by the isochrone goodness of fit to the $\pi_{12}$ calibrating subdwarf sample and provide the results in Table~\ref{table:ResultsHip}.  The distances found using $\pi_{12}$ calibration stars are on average 0.02 mag larger than when calibrating with our sample of $\pi_{10}$ subdwarfs, while the ages are 0.2 Gyr lower.  It should be noted that the most metal-poor calibration stars are those from \citet{Chab2016} and the \emph{HST} parallaxes were used both in this analysis and the previous analysis with the $\pi_{10}$ calibration sample as they are less uncertain.  Therefore, the difference between the GC distances and ages that are weighted by the fit to these calibration stars will be smaller.  If we remove these GCs from the comparison we find the $\pi_{12}$ distances and ages to be 0.04 mag larger and 0.3 Gyr younger than the $\pi_{10}$ weighted distances and ages, respectively. 

We find the overall uncertainties in distance modulus and age are the same when weighted by the $\pi_{10}$ and $\pi_{12}$ calibrating stars, as we are limited observationally by the reddening and in our models by the treatment of convection and diffusion along with uncertainties in the stellar opacities and nuclear reaction rates \citep{ChabSci,Grat2000,Grat2003}.  As mentioned previously, the 0.01 mag reddening uncertainty is propagated through our calculations using standard techniques and results in 0.03 mag uncertainty in GC distance modulus.  Additional photometric errors contribute 0.01 mag uncertainty as only cluster data with $\sigma_V<0.01$ were used to define the MS ridgeline.  The average total error we find for the GC distance modulus is 0.15 mag with the majority being contributed from the intrinsic uncertainties of the theoretical isochrones.  Similarly, for a given isochrone, we find the photometric errors and reddening uncertainty contribute 0.3 Gyr uncertainty in age.  Given that we find an average total age uncertainty of 1.6 Gyr, the majority of the uncertainty again stems from intrinsic uncertainties of the theoretical isochrones.

\begin{table*}[t]
\centering
\setlength{\tabcolsep}{4pt}
\caption{$\pi_{12}$ Weighted Distance Modulus and Age \label{table:ResultsHip}}
\begin{tabular}{rrrrrrrrr}
\hline\hline
& \multicolumn{2}{c}{ISO-P} && \multicolumn{2}{c}{ISO-VC} && \multicolumn{2}{c}{Combined}\\
\cline{2-3}\cline{5-6}\cline{8-9}
Cluster & $(m-M)_V$ & Age (Gyr) && $(m-M)_V$ & Age (Gyr) && $(m-M)_V$ & Age (Gyr)\\
\hline
NGC 104 & $13.55\pm0.20$ & $11.7\pm2.0$ && $13.53\pm0.20$ & $11.6\pm2.1$ && $13.54\pm0.20$ & $11.6\pm2.0$\\
NGC 288 & $14.97\pm0.15$ & $11.5\pm2.0$ && $15.01\pm0.15$ & $11.9\pm2.0$ && $15.00\pm0.15$ & $11.7\pm2.0$\\
NGC 2298 & $15.55\pm0.12$ & $13.3\pm1.4$ && $15.59\pm0.11$ & $13.6\pm1.5$ && $15.57\pm0.11$ & $13.4\pm1.5$\\
NGC 4590 & $15.40\pm0.10$ & $12.4\pm1.2$ && $15.43\pm0.10$ & $12.3\pm1.3$ && $15.41\pm0.10$ & $12.4\pm1.3$\\
NGC 5024 & $16.59\pm0.11$ & $13.0\pm1.3$ && $16.63\pm0.11$ & $12.6\pm1.3$ && $16.60\pm0.11$ & $12.8\pm1.3$\\
NGC 5053 & $16.32\pm0.11$ & $13.1\pm1.2$ && $16.36\pm0.10$ & $12.8\pm1.3$ && $16.34\pm0.10$ & $13.0\pm1.3$\\
NGC 5272 & $15.20\pm0.13$ & $10.9\pm1.7$ && $15.26\pm0.14$ & $10.2\pm1.8$ && $15.23\pm0.14$ & $10.5\pm1.8$\\
NGC 5466 & $16.15\pm0.11$ & $13.5\pm1.2$ && $16.18\pm0.10$ & $13.3\pm1.3$ && $16.16\pm0.10$ & $13.4\pm1.3$\\
NGC 5904 & $14.64\pm0.15$ & $9.8\pm2.0$ && $14.62\pm0.15$ & $10.2\pm2.0$ && $14.61\pm0.15$ & $10.0\pm2.0$\\
NGC 6101 & $16.07\pm0.11$ & $13.4\pm1.3$ && $16.07\pm0.11$ & $13.4\pm1.4$ && $16.07\pm0.11$ & $13.4\pm1.5$\\
NGC 6205 & $14.56\pm0.12$ & $12.0\pm1.5$ && $14.57\pm0.13$ & $11.9\pm1.7$ && $14.56\pm0.13$ & $12.0\pm1.7$\\
NGC 6341 & $14.79\pm0.11$ & $13.5\pm1.2$ && $14.82\pm0.12$ & $13.3\pm1.3$ && $14.80\pm0.12$ & $13.4\pm1.5$\\
NGC 6362 & $14.75\pm0.19$ & $11.5\pm2.0$ && $14.78\pm0.20$ & $11.2\pm2.1$ && $14.77\pm0.19$ & $11.3\pm2.0$\\
NGC 6541 & $15.02\pm0.12$ & $12.1\pm1.4$ && $15.02\pm0.12$ & $12.2\pm1.5$ && $15.02\pm0.12$ & $12.1\pm1.5$\\
NGC 6584 & $16.18\pm0.13$ & $11.4\pm1.5$ && $16.17\pm0.13$ & $11.5\pm1.7$ && $16.18\pm0.13$ & $11.4\pm1.7$\\
NGC 6652 & $15.26\pm0.19$ & $11.6\pm2.0$ && $15.32\pm0.19$ & $10.5\pm2.1$ && $15.30\pm0.19$ & $11.0\pm2.0$\\
NGC 6681 & $15.28\pm0.13$ & $12.5\pm1.5$ && $15.29\pm0.13$ & $12.4\pm1.7$ && $15.28\pm0.13$ & $12.5\pm1.7$\\
NGC 6723 & $14.85\pm0.20$ & $12.1\pm2.0$ && $14.91\pm0.20$ & $11.2\pm2.1$ && $14.88\pm0.20$ & $11.6\pm2.0$\\
NGC 6752 & $13.38\pm0.13$ & $12.5\pm1.6$ && $13.40\pm0.13$ & $12.3\pm1.7$ && $13.39\pm0.13$ & $12.4\pm1.7$\\
NGC 6809 & $14.20\pm0.12$ & $11.3\pm1.3$ && $14.20\pm0.11$ & $11.5\pm1.4$ && $14.20\pm0.11$ & $11.4\pm1.4$\\
NGC 7078 & $15.68\pm0.11$ & $12.6\pm1.2$ && $15.70\pm0.10$ & $12.4\pm1.3$ && $15.69\pm0.10$ & $12.5\pm1.3$\\
NGC 7099 & $14.86\pm0.11$ & $12.7\pm1.2$ && $14.89\pm0.10$ & $12.5\pm1.3$ && $14.87\pm0.11$ & $12.6\pm1.3$\\
\hline
\end{tabular}
\end{table*}

For many decades now, it has been known that the derived age for a given cluster depends on the prescribed $\alpha$-Fe ratio, mainly in the importance of oxygen and its role in the CNO-cycle \citep{SimIb1968,Salaris1993}.  Many groups have since attempted to quantify [$\alpha$/Fe] in both halo field and GC stars and typically find ranges near $\sim0.2$--0.5\,dex \citep{GratOrt1986,Barbuy1988,Sneden2004}. As the DSEP isochrones cover [$\alpha$/Fe] in steps of 0.2\,dex, we chose a binary distribution of [$\alpha$/Fe] = +0.2 and +0.4\,dex in our suite of models.  As expected, we find that the derived ages for each isochrone in the suite depend on the value of the [$\alpha$/Fe] construction parameter.

Although we do not find a large difference in the GC distance modulus determined using [$\alpha$/Fe] = +0.20\,dex isochrones versus those with [$\alpha$/Fe] = +0.40 dex, we do find a noticeable difference in age.  Specifically, we find that isochrones constructed using [$\alpha$/Fe] = +0.20\,dex give ages which are, on average, 0.6\,Gyr older than the mean GC age while isochrones constructed using [$\alpha$/Fe] = +0.40\,dex  give ages which are $\sim$0.4 Gyr younger.  It is important to note here that because we are calibrating these models along the MS where no difference is found between [$\alpha$/Fe] = +0.2 and +0.4\,dex models, both [$\alpha$/Fe] ratios give the same distributions of goodness of fit.

\begin{figure*}
\centering
\includegraphics[width=1.0\textwidth]{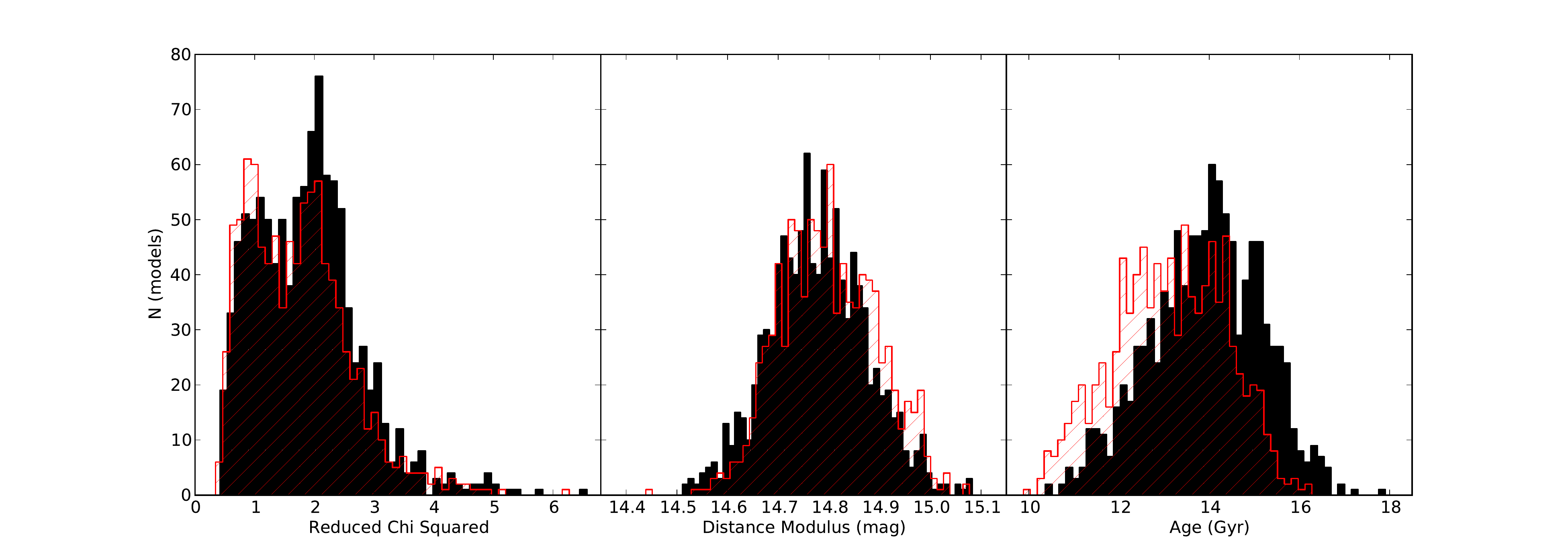}
\caption{\small{Distributions of $\chi_{\mathrm{Red}}^2$, M92 cluster distance and age for [$\alpha$/Fe] = +0.2\,dex (black) and +0.4\,dex(red) ISO-VC models.  \emph{Left} - Distribution of $\chi_{\mathrm{Red}}^2$ shows the overall goodness of fit to the calibration subdwarfs is relatively unaffected by [$\alpha$/Fe].  \emph{Center} - Distribution of M92 distances also shows that mean distance is unaffected by the choice of [$\alpha$/Fe].  \emph{Right} - However, the $\alpha$-Fe ratio has a significant impact on the cluster age, shifting the mean cluster age by 0.8 Gyr in the case of M92.}}
\label{fig:alpha}
\vspace{0.25cm}
\end{figure*}

We show this relationship Figure~\ref{fig:alpha} where in each panel the [$\alpha$/Fe] = +0.2\,dex ISO-VC models are in black while the [$\alpha$/Fe] = +0.40\,dex ISO-VC models are in red.  The left panel shows the distribution of $\chi_{\mathrm{Red}}^2$ based on the fit to the calibration stars.  The average $\chi_{\mathrm{Red}}^2$ for [$\alpha$/Fe] = +0.2 and +0.4\,dex are comparable, 1.87 and 1.79, respectively.  In the center panel we show the distribution of distance moduli of M92 obtained with the different [$\alpha$/Fe].  As one can see, the distance modulus is relatively unaffected with average distance modulus of  14.78 and 14.80\,mag for [$\alpha$/Fe] = +0.2 and +0.4.  On the other hand, the last panel shows how  [$\alpha$/Fe]  does impact the age of M92.  The [$\alpha$/Fe] = +0.2\,dex models give an average age of 13.9 Gyr, while the [$\alpha$/Fe] = +0.4\,dex models give an average age of 13.1 Gyr.

The difference in age predicted with the [$\alpha$/Fe] = +0.2 and +0.4\,dex models has been ascribed to the oxygen abundance and its importance in the CNO cycle.  Therefore, by having a higher [O/Fe] abundance, the CNO-cylce is able to progress easily with a lower temperature and power output, as evinced from the lower turn-off luminosity \citep[Figure 1]{VandenBerg1992}.  It is understandable then that the lower turn-off luminosity of an [$\alpha$/Fe] = +0.4\,dex model would in turn predict a lower age for a given GC.  It is also understandable that, given that the CNO-cycle dominates H-burning only from the turn-off onwards, we do not see a significant effect on the MS-fitting distances of the GCs between the [$\alpha$/Fe] = +0.2 and +0.4\,dex models.

By initially assuming that it is equally likely to find [$\alpha$/Fe] = +0.2 or +0.4\,dex in globular clusters, the quoted uncertainties in our derived ages take into account a systematic uncertainty of 0.2\,dex in [$\alpha$/Fe].  If one would like to extended the range of [$\alpha$/Fe] to include models of [$\alpha$/Fe]=+0.6\,dex, then the mean age for the most metal-poor clusters would decrease by 0.4\,Gyr, while the estimated uncertainty would increase by 0.2\,Gyr.

\section{Kinematical Study of GC Orbits}\label{Orbits}
Absolute proper motions are available for 17 GCs in our sample from the compilations and measurements of D. Casetti (2013)\footnote{http://www.astro.yale.edu/dana/gc.html}.  We supplemented this collection of proper motions with measurements for NGC\,5053, NGC\,6101, NGC\,6541 and NGC\,6652 from \citet{Khar2013} and measurements of NGC\,6681 from \citet{Massari2013} thus allowing for the calculation of full space velocities and Galactic orbits of the 22 clusters in our sample.  The cluster proper motions and radial velocities are provided in Table~\ref{table:Orbits} along with the cluster distance as found in this study.  These data will be combined with the cluster RA and DEC (Table~\ref{table:Clusters}) to be used as initial conditions in the orbit integration.

We model the Galactic orbits of our sample GCs using the \emph{galpy: A Python Library for Galactic Dynamics} \citep{Bovy2015}, specifically the orbit calculator with a Milky Way (MW)-like potential.  We use the built-in MWPotential2014 for this analysis which is fit to dynamical data of the MW as described in \citet{Bovy2015} in order to provide a realistic model on both small and large scales.  Briefly, this model is based on that of \citet{BovyRix2013} and combines an exponentially cutoff power-law density bulge (power-law exponent of -1.8 and cutoff radius of 1.9 kpc), a Miyamoto-Nagai disk, and a Navarro-Frenk-White dark matter halo.  We choose to use an updated solar distance from the Galactic center of $R_0 = 8.20\pm0.09$ kpc and circular velocity of $v_0 = 232.0\pm3.0$ km s$^{-1}$ from \citet{McMillan2016} and a total back integration time of 13 Gyr.

We provide the resulting orbital parameters in columns 7 -- 11 of Table~\ref{table:Orbits} with successive columns listing the perigalactic and apogalactic distance, the maximum distance from the Galactic plane, the orbital eccentricity and orbital energy.  In order to determine the uncertainty associated with each orbital parameter, we perform a systematic error analysis in which we vary each of the input variables ($\mu_\alpha\cos\delta$, $\mu_\delta$, $v_r$, $v_0$, and $R_0$) independently and calculate a new orbit for each cluster.  The uncertainties in the proper motions contribute the most to the uncertainties in the orbital parameters, typically changing the apogalactic  and perigalactic distances by a few percent to upwards of a factor of two.  On the other hand, the uncertainties associated with $v_0$ and $R_0$ contribute less than 5\% in most cases and, because the cluster radial velocities are known with such precision, they contribute less than 1\% to the uncertainty in $\mathrm{r}_\mathrm{apo}$ and $\mathrm{r}_\mathrm{peri}$ for the majority of the clusters.  The final uncertainty listed in Table~\ref{table:Orbits} is the combination of the individual effects added in quadrature.

\begin{turnpage}
\begin{table*}[t]
\centering
\setlength{\tabcolsep}{3pt}
\caption{Cluster Orbital Paramaters \label{table:Orbits}}
\begin{tabular}{l r r r r r r r r c r}
\hline\hline
Cluster & $\mu_\alpha\cos\delta$ & $\mu_\delta$ & $v_r$ & Distance & $R_{GC}$ & r$_\mathrm{peri}$ & r$_\mathrm{apo}$ & $z_\mathrm{max}$ & $e$ & E \\
 & (mas yr$^{-1}$) & (mas yr$^{-1})$ & (km s$^{-1}$) & (kpc) & (kpc) & (kpc) & (kpc) & (kpc) & & ($^*10^4$ km$^2/\mathrm{s}^2$)\\
\hline
NGC 104 & $5.63\pm0.21$ & $-2.73\pm0.29$ & $-18.00\pm0.10$ & $4.94\pm0.55$ & $7.14\pm0.01$ & $4.99^{+0.19}_{-0.16}$ & $7.64^{+0.07}_{-0.08}$ & $3.91^{+0.09}_{-0.08}$ & $0.21^{+0.02}_{-0.02}$ & $-5.80^{+0.16}_{-0.16}$\\
NGC 288 & $4.68\pm0.22$ & $-5.60\pm0.35$ & $-45.40\pm0.20$ & $9.46\pm0.86$ & $12.52\pm0.73$ & $3.92^{+0.72}_{-0.64}$ & $12.67^{+0.09}_{-0.09}$ & $10.31^{+0.32}_{-0.22}$ & $0.53^{+0.06}_{-0.06}$ & $-4.29^{+0.21}_{-0.19}$\\
NGC 2298 & $4.05\pm1.00$ & $-1.72\pm0.98$ & $148.90\pm1.20$ & $9.67\pm0.66$ & $15.04\pm0.42$ & $2.25^{+3.51}_{-2.06}$ & $16.34^{+1.08}_{-0.72}$ & $14.29^{+1.97}_{-7.69}$ & $0.76^{+0.21}_{-0.27}$ & $-3.56^{+0.63}_{-0.39}$\\
NGC 4590 & $-3.76\pm0.66$ & $1.79\pm0.62$ & $-94.70\pm0.20$ & $10.93\pm0.71$ & $8.34\pm0.52$ & $10.05^{+0.38}_{-0.51}$ & $42.30^{+13.59}_{-8.66}$ & $24.45^{+9.41}_{-7.79}$ & $-0.58^{+0.08}_{-0.12}$ & $0.06^{+0.86}_{-0.71}$\\
NGC 5024 & $0.50\pm1.00$ & $-0.10\pm1.00$ & $-62.90\pm0.30$ & $19.83\pm1.40$ & $21.22\pm1.32$ & $17.87^{+1.78}_{-4.63}$ & $44.78^{+96.50}_{-24.21}$ & $43.41^{+90.86}_{-30.49}$ & $0.43^{+0.38}_{-0.23}$ & $0.86^{+3.19}_{-1.87}$\\
NGC 5053 & $-5.81\pm0.53$ & $-2.76\pm0.53$ & $44.0\pm0.4$ & $17.77\pm1.17$ & $19.29\pm1.09$ & $15.86^{+2.02}_{-0.50}$ & $99.66^{+713.35}_{-111.56}$ & $98.13^{+125.51}_{-110.23}$ & $0.73^{+0.13}_{-0.15}$ & $2.85^{+1.96}_{-1.70}$\\
NGC 5272 & $-0.12\pm0.61$ & $-2.67\pm0.40$ & $-147.60\pm0.20$ & $10.71\pm0.91$ & $13.24\pm0.75$ & $4.54^{+1.14}_{-1.04}$ &$17.11^{+2.05}_{-1.38}$ & $14.60^{+1.55}_{-0.88}$ & $0.58^{+0.06}_{-0.06}$ & $-3.11^{+0.51}_{-0.42}$\\
NGC 5466 & $-3.90\pm1.00$ & $1.00\pm1.00$ & $110.70\pm0.20$ & $16.58\pm1.08$ & $17.90\pm0.98$ & $7.28^{+2.60}_{-2.53}$ & $98.79^{+211.71}_{-48.05}$ & $85.32^{+185.78}_{-47.87}$ & $0.83^{+0.11}_{-0.07}$ & $2.01^{+2.95}_{-2.02}$\\
NGC 5904 & $4.27\pm0.60$ & $-11.30\pm1.46$ & $53.20\pm0.40$ & $7.90\pm0.68$ & $8.23\pm0.30$ & $2.21^{+0.30}_{-0.13}$ & $47.13^{+29.06}_{-15.70}$ & $46.07^{+27.56}_{-15.91}$ & $0.91^{+0.04}_{-0.05}$ & $0.40^{+1.61}_{-1.34}$\\
NGC 6101 & $-2.98\pm1.62$ & $-0.46\pm1.62$ & $361.4\pm1.7$ & $14.12\pm0.97$ & $7.23\pm0.85$ & $5.24^{+4.19}_{-3.46}$ & $16.06^{+9.06}_{-1.10}$ & $8.72^{+11.80}_{-10.20}$ & $0.52^{+0.31}_{-0.14}$ & $-3.28^{+1.97}_{-0.35}$\\
NGC 6205 & $-0.10\pm0.80$ & $4.69\pm0.81$ & $-244.20\pm0.20$ & $7.90\pm0.62$ & $7.65\pm0.26$ & $6.81^{+0.40}_{-0.43}$ & $25.86^{+8.85}_{-6.20}$ & $25.36^{+8.67}_{-6.21}$ & $0.51^{+0.10}_{-0.09}$ & $-1.44^{+1.01}_{-0.88}$\\
NGC 6341 & $-3.58\pm0.89$ & $-0.60\pm0.60$ & $-120.00\pm0.10$ & $9.03\pm0.67$ & $10.39\pm0.36$ & $0.40^{+0.60}_{-0.31}$ &$10.85^{+0.48}_{-0.31}$ & $6.04^{+3.77}_{-0.73}$ & $0.93^{+0.06}_{-0.10}$ & $-5.58^{+0.27}_{-0.16}$\\
NGC 6362 & $-3.09\pm0.46$ & $-3.83\pm0.46$ & $-13.10\pm0.60$ & $8.10\pm0.88$ & $3.48\pm0.27$ & $2.14^{+0.15}_{-0.14}$ & $4.91^{+0.14}_{-0.13}$ & $2.60^{+0.13}_{-0.06}$ & $0.45^{+0.04}_{-0.03}$ & $-8.16^{+0.22}_{-0.21}$\\
NGC 6541 & $-3.24\pm1.04$ & $-1.74\pm1.04$ & $-158.7\pm2.3$ & $7.96\pm0.59$ & $2.23\pm0.15$ & $1.38^{+0.17}_{-0.20}$ & $3.81^{+0.72}_{-0.53}$ & $2.41^{+0.46}_{-0.30}$ & $0.47^{+0.09}_{-0.12}$ & $-9.93^{+0.73}_{-0.57}$\\
NGC 6584 & $-0.22\pm0.62$ & $-5.79\pm0.67$ & $222.90\pm15.00$ & $14.58\pm1.14$ & $7.73\pm1.02$ & $2.09^{+1.00}_{-0.91}$ & $16.62^{+5.05}_{-3.03}$ & $11.56^{+4.60}_{-3.04}$ &$0.78^{+0.07}_{-0.06}$ & $-3.55^{+1.17}_{-0.91}$\\
NGC 6652 & $5.25\pm$ 0.90& $-2.11\pm0.90$ & $-111.7\pm5.8$ & $9.68\pm1.06$ & $2.89\pm0.81$ & $1.66^{+0.10}_{-0.09}$ & $12.95^{+5.24}_{-3.24}$ & $8.97^{+4.30}_{-2.61}$ & $0.71^{+0.09}_{-0.10}$ & $-5.66^{+1.65}_{-1.42}$\\
NGC 6681 & $1.58\pm0.18$ & $-4.57\pm0.16$ & $220.30\pm0.20$ & $9.68\pm0.76$ & $3.10\pm0.54$ & $0.05^{+0.12}_{-0.08}$ & $6.39^{+0.23}_{-0.25}$ & $5.19^{+0.18}_{-0.18}$ & $0.98^{+0.02}_{-0.05}$ & $-7.88^{+0.33}_{-0.33}$\\
NGC 6723 & $-0.17\pm0.45$ & $-2.16\pm0.50$ & $-94.50\pm3.60$ & $7.46\pm0.85$ & $3.37\pm0.07$ & $1.46^{+0.09}_{-0.11}$ & $2.64^{+0.26}_{-0.22}$ & $2.37^{+0.17}_{-0.12}$ & $0.37^{+0.03}_{-0.05}$ & $-10.27^{+0.40}_{-0.37}$\\
NGC 6752 & $-0.69\pm0.42$ & $-2.85\pm0.45$ & $-26.70\pm0.20$ & $4.33\pm0.34$ & $5.32\pm0.13$ & $3.91^{+0.30}_{-0.16}$ & $5.50^{+0.03}_{-0.22}$ & $1.91^{+0.11}_{-0.01}$ & $0.17^{+0.01}_{-0.04}$ & $-7.44^{+0.24}_{-0.24}$\\
NGC 6809 & $-3.31\pm0.95$ & $-9.70\pm0.55$ & $174.70\pm0.30$ & $5.64\pm0.39$ & $4.58\pm0.10$ & $0.75^{+0.34}_{-0.28}$ & $5.91^{+0.34}_{-0.30}$ & $5.06^{+0.38}_{-0.32}$ & $0.77^{+0.09}_{-0.10}$ & $-8.02^{+0.32}_{-0.29}$\\
NGC 7078 & $-1.23\pm0.62$ & $-7.57\pm1.77$ & $-107.00\pm0.20$ & $11.74\pm0.76$ & $11.80\pm0.58$ & $10.27^{+0.27}_{-0.58}$ & $39.56^{+120.46}_{-23.98}$ & $29.16^{+97.30}_{-22.20}$ & $-0.55^{+0.29}_{-0.34}$ & $0.23^{+3.69}_{-2.62}$\\
NGC 7099 & $1.42\pm0.69$ & $-7.71\pm0.65$ & $-184.20\pm0.20$ & $8.20\pm0.62$ & $8.46\pm0.35$ & $3.16^{+0.89}_{-0.76}$ & $7.60^{+0.44}_{-0.34}$ & $6.26^{+0.39}_{-0.25}$ & $0.41^{+0.11}_{-0.10}$ & $-6.31^{+0.37}_{-0.31}$\\
\hline
\end{tabular}
\end{table*}
\end{turnpage}

\begin{figure*}
\centering
\includegraphics[width=1.0\textwidth]{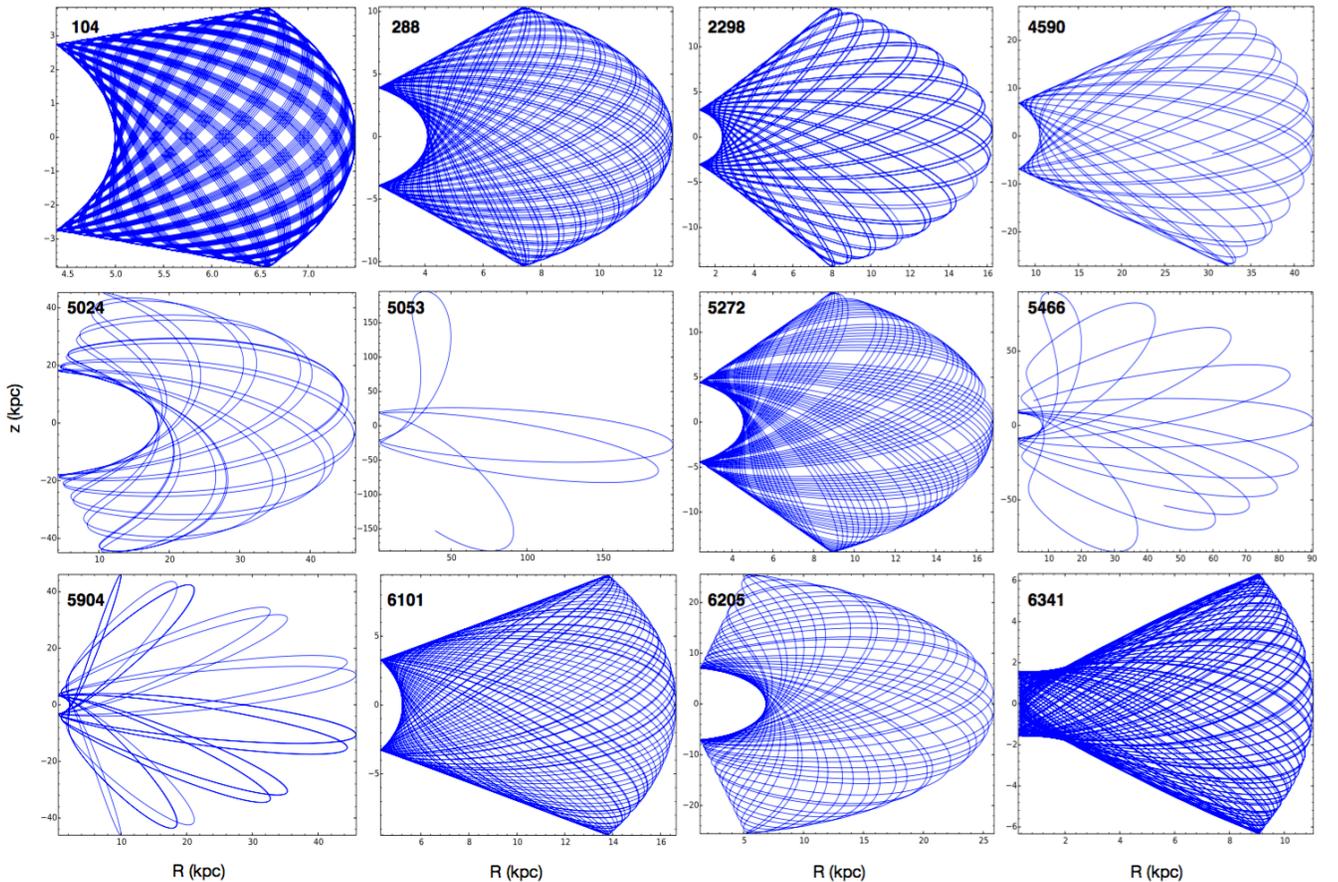}
\vspace{-1cm}
\caption{We show the meridional galactic orbits of twelve GCs integrated over the lifetime of the galaxy.  NGC\,104 (47 Tuc) is radially confined to within 8 kpc from the galactic center and vertically to only $\pm4$ kpc from the galactic plane and can therefore be considered an inner GC.  The other GCs have much more extend orbits and are therefore outer GCs. \label{fig:Orbits1}}
\vspace{0.25cm}
\end{figure*}

\begin{figure*}
\centering
\includegraphics[width=1.0\textwidth]{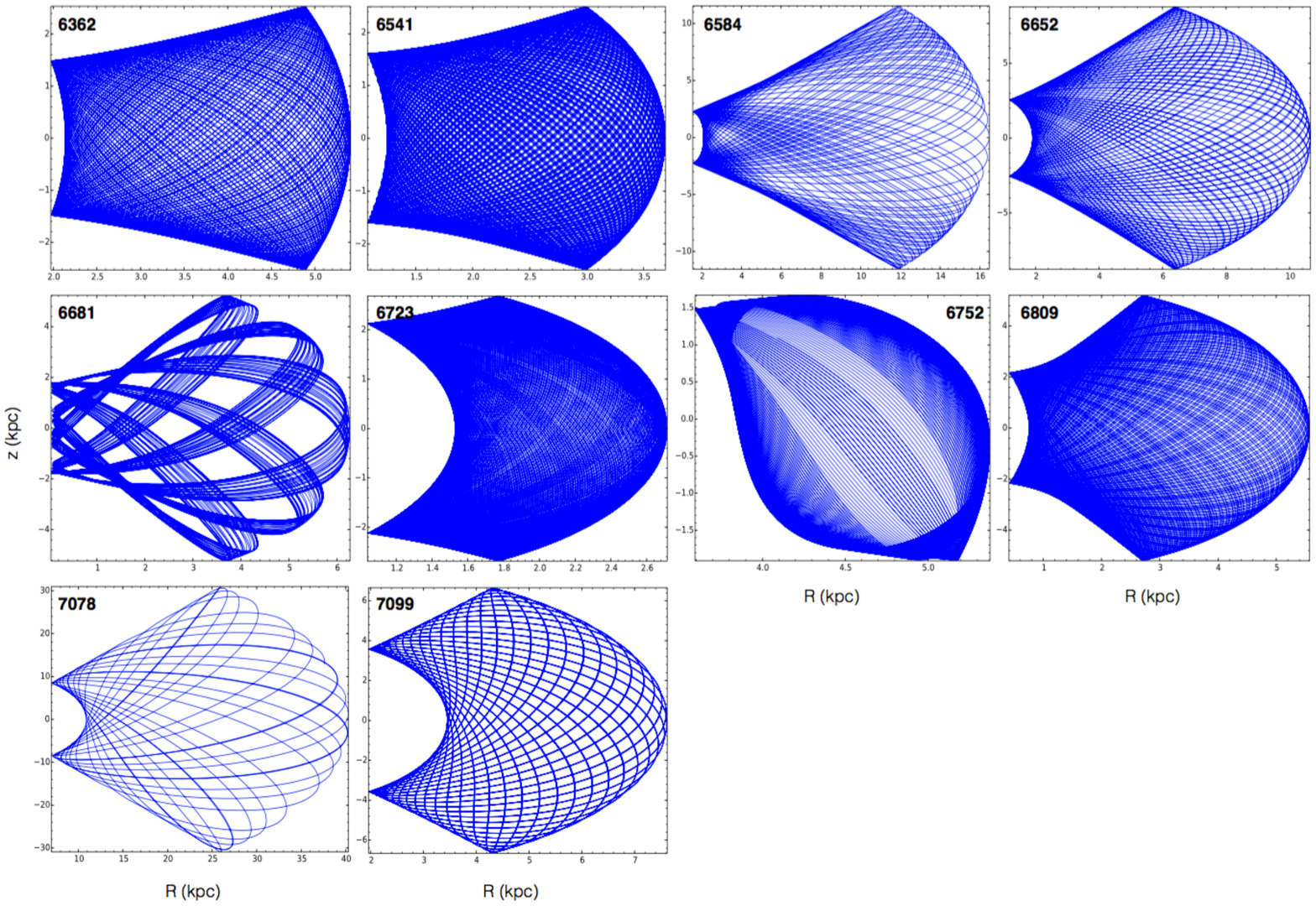}
\vspace{-1.5cm}
\caption{As in Figure~\ref{fig:Orbits1} we show meridional galactic orbits for the remaining 10 GCs in our sample.  Here NGC\,6362, NGC\,6541, NGC\,6723, NGC\,6752 and NGC\,6809 or considered inner GCs being associated with either the Galactic disk or bulge while the rest are outer GCs residing in the halo of the Galaxy. \label{fig:Orbits2}}
\vspace{0.25cm}
\end{figure*}

We show the meridional orbits for the clusters in Figures~\ref{fig:Orbits1} and \ref{fig:Orbits2} and find several different classes of GCs based on their orbital shapes.  Although the majority of the clusters in this sample reach sufficiently far distances from both the Galactic center and the Galactic plane to be considered outer halo GCs, we do find some that are associated with the Galactic disk and bulge.  The two clusters with the most confined orbits, NGC\,6541 and NGC\,6723, are characterized here as bulge GCs not extending farther than 4 kpc from the Galactic center either radially or vertically.  We find four additional disk clusters (47\,Tuc, NGC\,6362, NGC\,6681, NGC\,6752 and NGC\,6809) with slightly larger orbits, extending out to 8 kpc from the Galactic center radially, but confined to $\sim5$\,kpc from the plane. 

Previous studies have performed orbit integration using GC proper motion data \citep{Dinescu1999,Allen2006,Allen2008}; however, due to the differences in data sets and Galactic potential models, along with our updated GC distances it would be impractical to perform a detailed quantitative comparison of our resulting orbital parameters.  Nevertheless, we can still qualitatively consider how our results fit within the framework established by previous studies.  

In the study by \citet{Allen2006}, the authors perform a detailed orbital integration in both an axisymmetric MW-like potential and a non-axisymmetric barred potential.  By comparing the orbits of their 48 GCs in the axisymmetric potential versus those in the non-axisymmetric potential they found two groups of GCs, those whose pericenter distances are less than 4 kpc and whose orbits are affected by the presence of the bar (inner GCs) and those whose orbits are not (outer GCs) because they reside further than 4 kpc from the Galactic center.  Qualitatively, our results are in agreement as \citet{Allen2006} find the three clusters (NGC\,6362, NGC\,6723 and NGC\,6809) we identified as Disk/Bulge clusters to be inner GCs with orbits that are noticeably altered by the presence of the bar.  The clusters we characterize as halo clusters in this study are found to be outer GCs in \citet{Allen2006}.

\citet{Allen2006} also provide images of their meridional galactic orbits for 16 of their GCs, seven of which are in common with this study.  We find that five of the seven in common have the same general characteristics, covering the same range both radially and vertically.  The two discrepant GCs are NGC\,6362 which has an asymmetric shape in \citet{Allen2006} and NGC\,7078 which extends twice as far in z and almost four times as far in R in this study than in \citet{Allen2006}.  There are many factors that could contribute to these particular discrepancies including differences in the proper motions, positions, and distances to the clusters along with the choice of galactic potential.  For both NGC\,6362 and NGC\,7078 we use the same proper motion data and Galactic coordinates as in \citet{Allen2006}, therefore, we eliminate these as contributing factors.  \citet{Allen2006} do use a slightly larger solar Galactocentric distance, $R_0=8.5$ kpc, and a slightly lower circular velocity, $v_0=220$ km s$^{-1}$; however, we find no noticeable difference in the orbits of NGC\,6362 and NGC\,7078 when adopting these values.  In the case of NGC\,6362, the distance derived in this study ($d = 8.10\pm0.88$ kpc) is consistent with that of \citet{Allen2006} who found a distance of $7.6\pm0.8$ kpc.  On the other hand, we find a distance for NGC\,7078 ($d=11.74\pm0.76$ kpc) that is 30\% larger than that of \citet{Allen2006}.  In adopting the \citet{Allen2006} distance to NGC\,7078, we find the apogalactic distance to be significantly reduced, a mere 4 kpc larger than that of \citet{Allen2006} rather than four times the distance.  As is the case with the other clusters in the sample, the remaining small differences can be attributed to differences in the Galactic potential.

\citet{Dinescu1999} perform a similar orbital analysis with 16 of the 22 clusters in our sample in common.  Although the authors do not provide meridional orbits for comparison, we can examine their orbital parameters.  If we want to characterize the location of each GC in its orbit we can look at how near the cluster's $R_{GC}$ is to its $R_{apo}$.  In this case we will define``near apogalactic" as $R_{GC}/R_{apo}\geq0.70$, ``near perigalactic" as $R_{GC}/R_{apo}\leq0.30$ and ``mid-orbit" as anything else.  Of the 16 clusters in common with \citet{Dinescu1999} we find 8 to be near apogalactic, 3 near perigalactic, and 5 in mid-orbit.  On the other hand, the results from \citet{Dinescu1999} suggest that 9 clusters are near apogalactic, 5 in mid-orbit and only 2 near perigalactic.  The  clusters that are in disagreement in this comparison are NGC\,4590, NGC\,6809 and NGC\,7078.  In the characterization of NGC\,4590 using the results of \citet{Dinescu1999} the cluster was considered to be mid-orbit with $R_{GC}/R_{apo}=0.38$ whereas we classified the cluster as being near perigalactic with $R_{GC}/R_{apo}$ at 0.20.  This is similar to the difference seen for NGC\,6809 in which \citet{Dinescu1999} finds $R_{GC}/R_{apo}=0.68$ where we find $R_{GC}/R_{apo}=0.78$.  The discrepancies can be attributed to the difference in galactic potentials and underlying uncertainty in the galactic orbit parameters.  The discrepancy was significantly larger for NGC\,7078 where the results of \citet{Dinescu1999} suggest the cluster is near apogalactic with $R_{GC}/R_{apo}=0.95$ and yet we find it to be near perigalactic with $R_{GC}/R_{apo}=0.30$.  This difference is largely due to the difference in apogalactic distance which, as we saw in our comparison to \citet{Allen2006}, is a direct result of a larger derived distance and differences in the Galactic potential.  However, it is important to note that given the large uncertainty associated with the apogalactic distance, it is hard to say with certainty exactly where this cluster is in its orbital trajectory.

\section{Age-[Fe/H] and Age-R$_{GC}$ Relations}\label{AgeRel}
As the ages of GCs have long been used to glean information about the origins of our Galaxy it is important to consider how the ages we determined here correlate with other important properties and what that means in terms of our theories of galactic formation and evolution.  Calculating the orbits of these GCs allows us to study where they may have originated and better what their origin means for the formation and evolution of our Galaxy.  Along with their origins, though, we are also interested to uncover any underlying relationship between the inherent characteristics of GCs such as their age, metallicity and current and/or past locations in the Galaxy.   

\begin{figure*}
\centering
\includegraphics[width=1.0\textwidth]{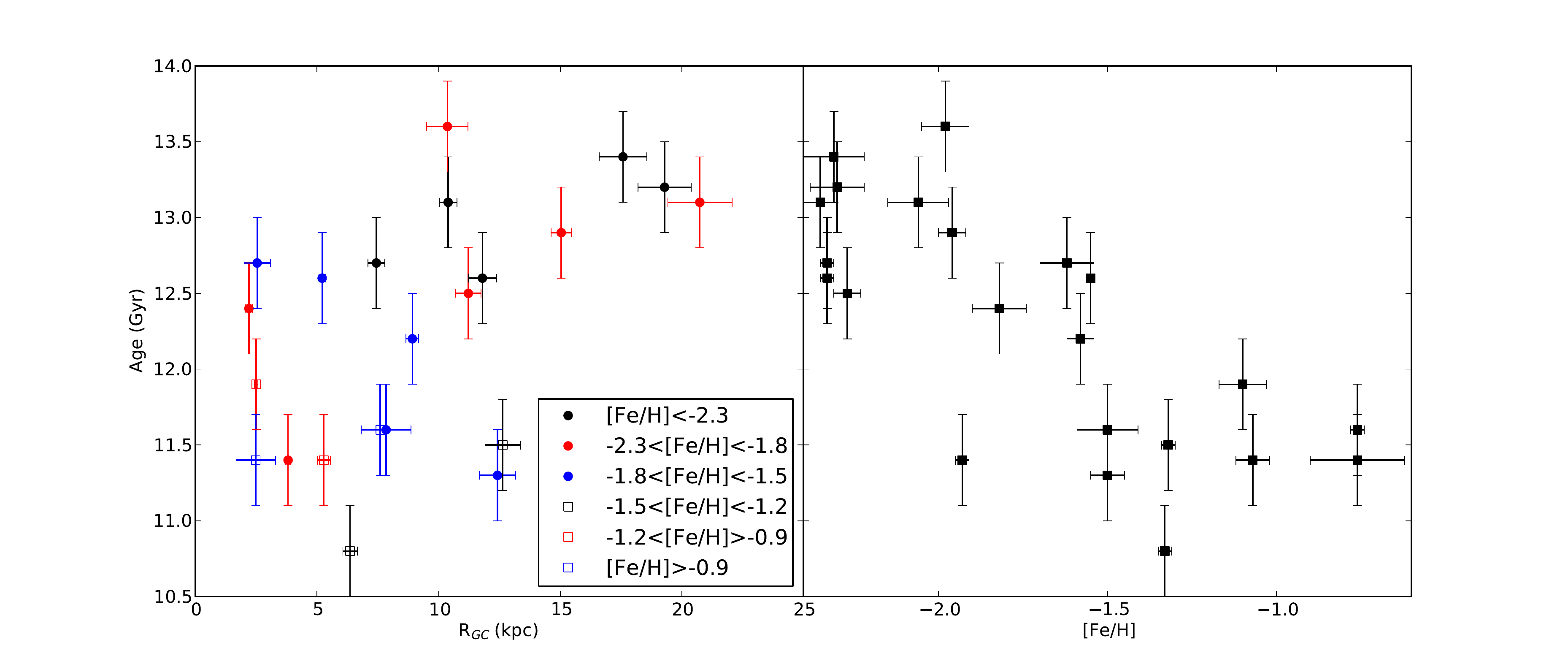}
\caption{\emph{Left} Age is shown as a function of Galactocentric distance with different metallicity GCs marked as shown.  We do not find strong evidence for any relation between age and $R_{GC}$.  \emph{Right} -- The age-metallicity relation is shown for the 22 Milky Way GCs in this sample.  The point markers show the age and metallicity of the individual GCs while the horizontal and vertical errorbars represent the [Fe/H] uncertainty as determined by \citet{Carr2009} and the 0.3 Gyr random error in the age estimates. \label{fig:AgeMet}}
\vspace{0.5cm}
\end{figure*}

In the left panel of Figure~\ref{fig:AgeMet} we show GC age as a function of Galactocentric distance ($R_{GC}$).  Although there is no clear dependence of age on $R_{GC}$ we do find that the most metal-poor GCs ([Fe/H]$<$-1.8 dex) are located over the full range of $R_{GC}$, but the same cannot be said for the more metal-rich GCs which are clustered within $R_{GC}<15$ kpc.  This agrees with that of \citet{VandenBerg2013} who find no trend in age or age dispersions with $R_{GC}$ for the set of 55 GCs in their study.  \citet{MF2009} perform a similar analysis using relative ages determined with two different sets of isochrones and in fact find a trend in age with galactocentric distance in both cases.  This would seem to be at odds with our finding of no age-$R_{GC}$ trend; however, it is important to note that \citet{MF2009} consider GCs with [Fe/H] ranging from -2.02 dex to -0.18 dex, extending to much more metal-rich clusters than we consider here.  If we only look at their clusters with [Fe/H]$<-0.8$ dex, then their trend becomes negligible and we find agreement in our results.

We also investigate the well-studied age-metallicity relation in the right panel of Figure~\ref{fig:AgeMet}.  Previous studies found not only a simple trend of metal-poor GCs being among the oldest \citep{C1996}, but also a visible bifurcation in the age-[Fe/H] relation \citep{MF2009,Dotter2010,Dotter2011,Leaman2013,VandenBerg2013}.  Although we do not find the dramatic bifurcation of \citet{Leaman2013} and \citet{VandenBerg2013}, we do find that our results are at least consistent.  The results shown here are more compatible with those of \citet{MF2009} who find an age-[Fe/H] relation with two branches: one having a clear trend of decreasing age with increasing metallicity for the young clusters while the other is a mostly coeval old branch that has a larger dispersion.  As the studies of \citet{MF2009}, \citet{Leaman2013} and \citet{VandenBerg2013} use more than twice as many clusters, they are able to identify the relationship between age and metallicity with much more confidence.  Although our results are not in obvious disagreement with the previous studies, we are limited in the amount of additional evidence our study can supply in support of a bifurcated age-metallicity relation due to the number of clusters studied and the larger scatter in our results.

A bifurcated age-metallicity relation would have strong implications for how we understand the formation of the Milky Way.  \citet{Leaman2013} suggest that the offset seen between the two sequences offers evidence that the higher metallicity disk sequence clusters form in situ within in the disk and that these clusters are the most metal-rich clusters at any given age.  They also note that the offset suggests that the lower metallicity clusters are remnants from hosts the size of the SMC, WLM, and maybe even larger satellite galaxies such as the LMC and Sagittarius.  

\section{Summary}\label{Summary}
We use the methods of \citet{Chab2016} to determine the MS-fitting distances and ages of 22 Milky Way GCs from the \emph{HST} ACS GC Treasury Project \citep{Sara2007} spanning a range of metallicities, $-2.4\leq[Fe/H]\leq-0.7$ dex, and reddening, $E(B-V)\leq0.22$ mag.  Both TGAS and \emph{HST} parallaxes of subdwarfs with metallicities between -2.7 dex and -0.6 dex are used to calibrate the stellar evolution models used in the MS-fitting.  As in \citet{Chab2016}, a MC method is used to construct isochrones with input parameters that cover the parameter space.  In doing so, it is possible to not only determine the MS-fitting distances and ages of the GCs, but also the combined observational and theoretical uncertainty.

The distance modulus and corresponding uncertainty for each of the clusters is an average of the distance found using each of the isochrones weighted by the isochrone goodness of fit to the calibration subdwarfs.  We find an average distance modulus uncertainty of 0.13 mag along with an offset between the ISO-P and ISO-VC distances that increased with decreasing metallicity and has a direct impact on the ages, resulting in an average offset of $\pm0.2$\,Gyr between ISO-P and ISO-VC ages for a given cluster with ISO-P ages being larger for GC with [Fe/H]$<-1.5$\,dex and visa versa with [Fe/H]$>-1.5$\,dex.

The uncertainty in our distance and age estimates for these GCs incorporates the observational uncertainty associated with the photometry and reddening estimates while also taking into account the theoretical uncertainty associated with the stellar evolution models.  As was shown in \citet{Chab2016}, this theoretical uncertainty is most strongly attributed to the choice of mixing length parameter, diffusion coefficients, atmospheric $T(\tau)$ relation, pp-chain reaction rate and high temperature opacity. 

With updated distances and ages for these clusters, along with proper motion measurements from previous studies, we are able to calculate orbits for these clusters and infer information about their origins and the formation of our Galaxy.  We find the majority of the clusters to be outer halo clusters, with a handful being confined to the Galactic bulge or disk.  We do not find any correlation between the GC Galactocentric distance and the cluster's age; however, we do find a an age-metallicity relation that is roughly compatible with the bifurcated relations of \citet{MF2009}, \citet{Leaman2013} and \citet{VandenBerg2013}.  As some of the clusters are very loosely bound with orbits extending to distances at which interactions with nearby dwarf galaxies could occur, it would be interesting to study the GC populations of dwarf galaxies such as the Large and Small Magellanic Clouds to determine if similar patterns exist in those environments, thereby providing further evidence for the origins of MW GCs.  

\acknowledgements
The authors thank the anonymous referee for giving constructive scientific comments and suggestions to improve the manuscript.  This material is based upon work supported by the National Science Foundation Graduate Research Fellowship under Grant No. DGE-1313911. Any opinion, findings, and conclusions or recommendations expressed in this material are those of the authors(s) and do not necessarily reflect the views of the National Science Foundation. This work is supported in part by grant AST-1211384 from the National Science Foundation.  This work has made use of data from the European Space Agency (ESA) mission \emph{Gaia} (\url{http://www.cosmos.esa.int/gaia}), processed by the \emph{Gaia} Data Processing and Analysis Consortium (DPAC, \url{http://www.cosmos.esa.int/web/gaia/dpac/consortium}). Funding for the DPAC has been provided by national institutions, in particular the institutions participating in the {\it Gaia} Multilateral Agreement.  This work made use of NASA's Astrophysics Data System (ADS) and the SIMBAD database, operated at CDS, Strasbourg, France.

\clearpage
\appendix
\section{Lutz-Kelker Correction}
One bias that may be present in our sample of calibration subdwarfs is the classical Lutz-Kelker \citep{LK1973} correction. When a star has a measured parallax value higher than its true value it is more likely to be included in a sample than if the parallax measured was lower than its true value.  Additionally, since we require that $\sigma_{\pi}/\pi\leq0.1$, stars with a higher measured parallax will therefore be weighted more strongly. 

To test for this bias in our sample we created a Monte Carlo simulation following that of \citet{Chab1998}. Each source of possible uncertainty either due to instrumental effects, selection effects, or number statistics were modeled and included in the simulation.  Below we provide the steps used to generate synthetic stars that closely resemble our actual data.

\begin{enumerate}
\item A non-normalized probability function of [Fe/H] was derived for all the stars that were in the Tycho-2 and Stromgren catalogs using only references more recent than the year 2000.

$P([\textrm{Fe/H}])=\exp{(5.19227+1.26595[\textrm{Fe/H}])}$
\item The [Fe/H] probability function is normalized and used to compute a random list of true [Fe/H] values, [Fe/H]$_{t}$, for the stars. 
\item Gaussian uncertainties in [Fe/H] are randomly selected with $\sigma_{\textrm{[Fe/H]}}$=0.10 and 0.15.  This yields an observed [Fe/H], [Fe/H]$_{o}$.
\item Random distances are chosen assuming a sphere of constant stellar density out to 400 pc in steps of 10 pc.  Using the computed distances, an actual parallax, $\pi_{t}$, is created for each star. 
\item Masses are included using a Salpeter initial mass function from 0.4 -- 0.9 $M_{\odot}$.
\item A mass-luminosity function for MS stars was found using DSEP isochrones. The following fits were used, interpolating between functions based on [Fe/H] values.

$M_V = 15.011-11.8925$ m \hspace{1.2cm}([Fe/H] $>$ -0.5)

$M_V = 14.6153 - 12.5644$ m \hspace{1cm}(-0.5 $>$ Fe/H] $>$ -1.0)

$M_V = 14.2013 - 12.6459$ m \hspace{1cm}(-1.0 $>$ [Fe/H] $>$ -1.5)

$M_V = 14.0212 - 12.7825$ m \hspace{1cm}(-1.5 $>$ [Fe/H] $>$ -2.0)

$M_V = 13.5781 - 12.1817$ m \hspace{1cm}(-2.0 $>$ [Fe/H] $>$ -2.5)

$M_V = 16.6547 - 16.5498$ m \hspace{1cm}(-2.5 $>$ [Fe/H] $>$ -3.0)

$M_V = 16.7498 - 16.6888$ m \hspace{1cm}(-3.0 $>$ [Fe/H] $>$ -3.5)

$M_V = 16.0395 - 15.6256$ m \hspace{1cm}(-3.5 $>$ [Fe/H])
\item Using $M_{V}$ and $\pi_{t}$, a true apparent magnitude, $V_{t}$, was found. 
\item A random Gaussian uncertainty in absolute magnitude, $\sigma_{m_{V}}= 0.02$, was added to the actual magnitude to create an observed apparent magnitude $V_{o}$.
\item \emph{Gaia} completeness -- the TGAS stars were compared to the stars included in \emph{Gaia} Data Release 1 as a function of magnitude, giving a probability of inclusion as a function of magnitude. 
\item Uncertainties for the TGAS parallax were handled two different ways.  Both methods produced the same Lutz-Kelker corrections.
\begin{itemize}
\item Gaussian uncertainties for parallax are computed using a fit of parallax uncertainties versus magnitude from \citet{MLH2015} and used to find the observed parallax, $\pi_{o}$.

$\sigma_{\pi}=7.56106-2.3672V_{to}+0.248459V_{to}^{2}-0.00822623V_{to}^{3}$ (mas)
\item Gaussian uncertainties for parallax are computed using Table 1 from \citet{MLH2015}. and used to find the observed parallax, $\pi_{o}$.
\end{itemize}
\item The observed absolute magnitude, $M_{Vo}$, is calculated as follows:
$M_{Vo}=V_{o}+5.0\log(\pi_{o})+5.0$
\item Synthetic stars are accepted into the sample if the following are true:  $M_{Vo} > 5.5$, $\sigma_{\pi}/\pi_{o} < 0.10$ and $[\textrm{Fe/H}]_{o} < -0.60$.
\end{enumerate}

A sample of $10^7$ synthetic stars was created by the simulation with varying amounts of uncertainty included.  A synthetic star is accepted if the simulation finds that could be observed; in this case, $10^4$ synthetic stars are accepted.  Based on these $10^4$ stars, we find no appreciable Lutz-Kelker correction.  

The bias we find in absolute magnitude is at most $M_{V}=+0.009$ mag and in metallicity of at most [Fe/H]=-0.171 dex.  The [Fe/H] bias may seem quite large at first glance, but it is important to examine the bias in each individual metallicity bin.  Table~\ref{table:LKbias} gives the bias in absolute magnitude and [Fe/H] in each metallicity bin.  The largest [Fe/H] bias in an individual bin was [Fe/H]=-0.004 dex with most bins below [Fe/H]=-0.001 dex.

\begin{table*}[t]
\centering
\caption{Metallicity Binned Lutz-Kelker Bias \label{table:LKbias}}
\begin{tabular}{l | cc | cc}
\hline\hline
&  \multicolumn{2}{c}{$\sigma_{[Fe/H]} = 0.10$}\vline & \multicolumn{2}{c}{$\sigma_{[Fe/H]} = 0.15$}\\
\cline{1-3}\cline{4-5}
[Fe/H] bin & $M_V$ bias (mag) & [Fe/H] bias (dex) & $M_V$ bias (mag) & [Fe/H] bias (dex) \\
\hline
-0.77 & 0.006  & 0.001 & 0.009  & 0.002 \\
-1.17 & -0.003 & -0.004 & 0.008  & 0.002 \\
-1.33 & 0.009  & -0.001 & -0.004 & -0.002 \\
-1.81 & 0.005  & -0.003 & -0.006 & -0.001 \\
-2.00 & -0.004 & -0.001 & -0.008 & -0.001 \\
-2.43 & 0.009  & -0.002 & 0.005  & -0.002 \\
\hline
\end{tabular}
\vspace{0.25cm}
\end{table*}

\clearpage

\end{document}